\documentclass[manuscript]{acmart}
\AtBeginDocument{%
  \providecommand\BibTeX{{%
    \normalfont B\kern-0.5em{\scshape i\kern-0.25em b}\kern-0.8em\TeX}}}

\setcopyright{acmcopyright}
\copyrightyear{2018}
\acmYear{2018}
\acmDOI{XXXXXXX.XXXXXXX}

\acmJournal{JACM}
\acmVolume{37}
\acmNumber{4}
\acmArticle{111}
\acmMonth{8}

\usepackage{tabularx, multirow}
\usepackage{pifont}
\usepackage{xcolor,colortbl}
\usepackage{fontawesome5}
\begin{document}

\newcommand{\rqone}{What are the tools and frameworks proposed to generate and configure access control policies?}

\newcommand{\rqtwo}{What are the limitations of the existing tools and framework developed to generate and configure access control policies?}

\newcommand{\pc}{49 }

\title{SoK: Access Control Policy Generation from High-level Natural Language Requirements}

% \title{SoK: Towards Access Control Policy Generation from Natural Language}

% \title{SoK: Towards Access Control Policy Configuration using Natural Language Requirements}


\author{Sakuna Harinda Jayasundara}
\email{sjay950@aucklanduni.ac.nz}
\orcid{0009-0006-7932-5204}
\affiliation{%
  \institution{University of Auckland}
  \city{Auckland}
  \country{New Zealand}
}
\author{Nalin Asanka Gamagedara Arachchilage}
% \email{nalin.arachchilage@auckland.ac.nz}
\email{nalin.arachchilage@gmail.com}
\orcid{0000-0002-0059-0376}
\affiliation{%
  \institution{University of Auckland}
  \city{Auckland}
  \country{New Zealand}
}

\author{Giovanni Russello}
\email{g.russello@auckland.ac.nz}
\orcid{0000-0001-6987-0803}
\affiliation{%
  \institution{University of Auckland}
  \city{Auckland}
  \country{New Zealand}
}

% \affiliation{%

% \affiliation{%

% \affiliation{%

% \affiliation{%

% \affiliation{%

% \affiliation{%

% \affiliation{%

\renewcommand{\shortauthors}{Jayasundara et al.}

\begin{abstract}
Administrator-centered access control failures can cause data breaches, putting organizations at risk of financial loss and reputation damage. Existing graphical policy configuration tools and automated policy generation frameworks attempt to help administrators configure and generate access control policies by avoiding such failures. However, graphical policy configuration tools are prone to human errors, making them unusable. On the other hand, automated policy generation frameworks are prone to erroneous predictions, making them unreliable. Therefore, to find ways to improve their usability and reliability, we conducted a Systematic Literature Review analyzing 49 publications, to identify those tools, frameworks, and their limitations. Identifying those limitations will help develop effective access control policy generation solutions while avoiding access control failures.
\end{abstract}





\maketitle

\section{Introduction}
\label{sec:intro}

In June 2023, terabytes of Microsoft's sensitive information were exposed through a GitHub repository due to an access control failure \cite{Page_2023}. Microsoft AI (Artificial Intelligence) researchers published a GitHub repository in June 2023 as a part of their ongoing research, allowing the users to download open-source code and AI models for image recognition using an Azure storage URL (Uniform Resource Locator) \cite{Page_2023}. However, instead of allowing users only to access and read the source code and AI models in a specific storage bucket, the storage account administrator has accidentally given "full access" to the entire storage account through that URL \cite{Page_2023}. As a result, 38 terabytes of sensitive information, including passwords to Microsoft services, private keys, personal backups of two company employees, and more than 30,000 Microsoft Teams messages of hundreds of employees, stored in the storage account, were exposed to the public \cite{Page_2023}. It allowed anyone not only to read but also to change that private information, making the entire organization and its employees victims of a data leak. 
This incident shows how severe the mistakes of an administrator can be when it comes to access control. 

Therefore, to avoid such administrator-centered access control failures, previous literature proposed graphical policy configuration (i.e., policy authoring and visualization) tools that guide the administrators to write and visualize policies manually from high-level access control requirements without worrying about complex access control languages and their syntax \cite{stepien2009non,stepien2014non, bertard2020using, turner2017proposed, brostoff2005r, zurko1999user, johnson2010usable, johnson2010optimizing, maxion2005improving, cao2006intentional, reeder2007usability, reeder2008expandable, morisset2018visabac, reeder2011more, karat2006evaluating, morisset2018building}. However, manual policy authoring is a repetitive, laborious, and error-prone task \cite{narouei2017identification, nobi2022machine}. The administrator has to repetitively write policies one by one so that the appropriate access is provided to correct resources \cite{reeder2008expandable, reeder2011more, maxion2005improving}. For example, when administrators have several natural language access requirements to apply to the authorization system, they have to go through the requirements one by one manually, first to identify the underlying rules, secondly, to identify the policy components (e.g., users, actions, and resources) of the rules and finally use those components to build the policy using the graphical policy authoring tool. This manual process becomes even harder when there are usability issues of the configuration interface that induce human errors \cite{reeder2007usability}. 
Therefore, many administrators consider manual access control policy configuration as an overhead that makes them burned out and stressed, which leads to accidental human errors \cite{botta2007towards, palmer_2021}. As a solution, researchers then developed fully-automated policy generation frameworks to remove the system administrator almost entirely from policy generation \cite{shi2011controlled, alohaly2018deep, alohaly2021hybrid, fatema2016semi, slankas2013access, heaps2021access, slankas2013accessid, brodie2006empirical, xia2022automated, alohaly2019automated, xiao2012automated, narouei2018automatic, narouei2015automatic, slankas2012classifying, singru2020efficient, vaniea2008evaluating, rosa2020parser, inglesant2008expressions, liu2017automated, narouei2017identification, ye2016identify, perry2013modular, basile2010ontology, yang2021purext, slankas2014relation, narouei2017towards, alohaly2019towards, narouei2015towards, tanoli2018towards, zhu2021unstructured, brodie2005usable, shen2021using, abdelgawad2023synthesizing}.
Those frameworks translate the natural language access control policies (NLACPs) in high-level requirement specification documents into machine-executable policies automatically using natural language processing (NLP) and machine learning (ML) techniques \cite{narouei2017identification}. Therefore, the stress and fatigue due to the policy engineering overhead will be alleviated. However, the existing automated solutions are not reliable enough to generate access control policies without being verified by a human expert \cite{kaur2021human, del2022systematic}, as the ML/NLP techniques used to develop those frameworks do not produce accurate results always \cite{kaur2021human}, as they are often prone to a significant number of false positives and false negatives \cite{del2022systematic}.

In summary, even though graphical policy authoring and visualization tools attempt to guide the administrator in writing and visualizing policies even without the knowledge about access control languages, their own limitations induce human errors, making them \textbf{unusable} for accurate policy authoring and visualization. On the other hand, NLP-based automated policy generation frameworks are \textbf{not reliable enough} to generate machine-executable access control policies accurately without human supervision. Therefore, to help the administrator avoid access control failures due to such challenges, it is crucial to identify ways to improve the usability and reliability (i.e., address the usability-security trade-off) of those tools and frameworks. 
To do that, first, we have to identify what are the existing policy configuration (i.e., policy authoring and visualization) and generation tools and frameworks and their limitations. With that in mind, we conduct this Systematic Literature Review (SLR) to answer the following key research questions:

\textbf{RQ1}: \rqone

\textbf{RQ2}: \rqtwo

The rest of the article is organized as follows. In Section \ref{sec:related_works}, we briefly report related works to this SLR and highlight the research gap and the contribution to knowledge. Then, in Section \ref{sec:slr}, we discuss the methodology we followed to plan and conduct the SLR, followed by the results in Section \ref{sec: results}. After reporting results, we discuss them and provide guidelines to further improve the identified tools and frameworks in Section \ref{sec:discussion}. Finally, we discuss some of the limitations of this SLR in Section \ref{sec:threats} followed by conclusions and future works in Section \ref{sec:conclusion}.

% We conducted this study by thematically analyzing 53 publications related to top-down access control policy generation. We followed the methodology proposed by Kitchenham \cite{kitchenham2004procedures} and presented results according to the PRISMA framework \cite{page2021prisma} under 2 main themes: GUI-based policy authoring and visualization, and NLP-based policy generation. Under GUI-based policy authoring and visualization, we report results in terms of text editor-based, template-based, access matrix-based, and graph-based tools. Furthermore, under the NLP-based policy generation, we analyze existing NLP policy generation pipelines by breaking them down into 5 steps: preprocessing, text classification, information extraction, information transformation, and evaluation. Finally, we report their limitations and provide guidelines to improve them further. 

\section{Related Work}
\label{sec:related_works}

\begin{figure*}[]
    \centering
    \includegraphics[height=3.8cm]{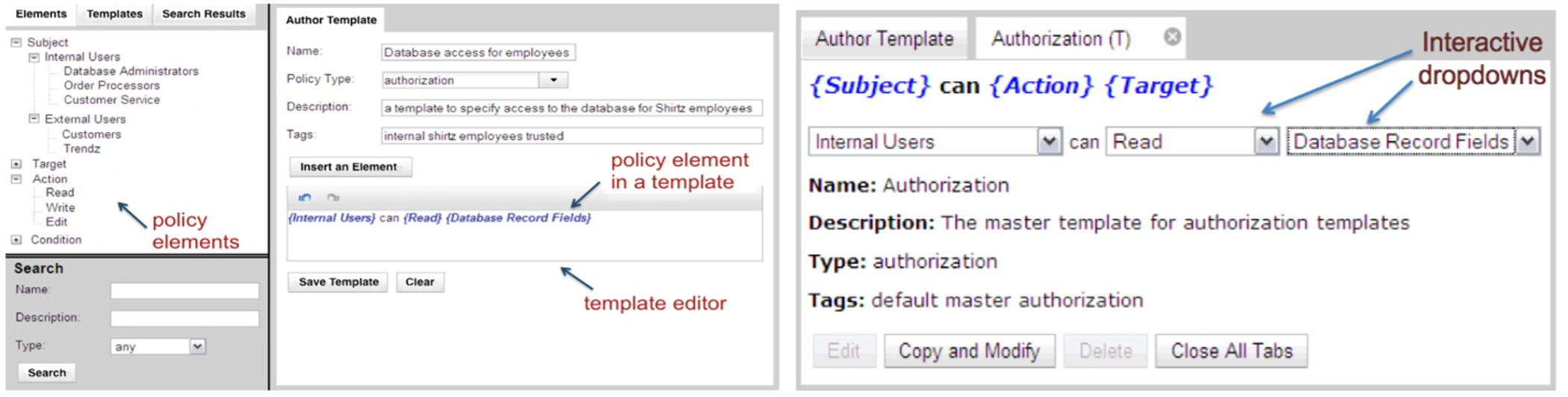}
    \Description[A template-based policy authoring interface]{The image shows a policy authoring interface that provides templates with placeholders to input policy components and build policies.}
    \caption{A template-based policy authoring interface (left) and its template designer (right) that defines templates \cite{johnson2010optimizing, johnson2010usable}.}
    
    \label{fig:template}
\end{figure*}

Previous research initially attempted to solve access control failures by guiding the system administrator to write policies via graphical user interfaces (GUIs) \cite{reeder2008expandable, morisset2018visabac}. Consider a template-based policy authoring interface as shown in Fig.\ref{fig:template}. It provides pre-defined templates such as \texttt{\{Subject\} can \{Action\} \{Target\}} to write policies by choosing suitable policy components for the placeholders denoted between "\{\}" \cite{johnson2010usable, johnson2010optimizing}. This interface points out what the necessary policy components are (i.e., subject, action, and target) and in what way those components should be organized in the policy \cite{johnson2010usable}. Therefore, by following the provided template, administrators can avoid incorrect access control policies (i.e., policies containing wrong policy components in wrong placeholders) and incomplete access control policies (i.e., policies without necessary policy components). However, the usability issues of the existing tools make policy authoring and visualization difficult for administrators \cite{inglesant2008expressions}. For example, sometimes, their intended policies cannot be easily written using the provided policy template. Assume that the administrator has to write a policy, \emph{"Bob is allowed to access the computer "A" if the time is between 9 a.m. and 5 p.m."}, using the mentioned template-based authoring interface. If the interface does not support conditions (\emph{"if the time is between 9 a.m. and 5 p.m."}), the administrator might neglect the condition entirely and generate a policy that allows Bob to access the computer at all times, without any restrictions. These situations may result in access control failures leading to data breaches \cite{reeder2007usability, maxion2005improving, brostoff2005r}. To avoid such situations, it is imperative to understand where these limitations are in graphical policy authoring and visualization tools and provide solutions. 
With that motivation, previous literature that proposed such tools evaluated their own tools individually and pointed out their unique usability issues via user studies \cite{reeder2008expandable, reeder2011more, reeder2007usability, johnson2010usable}. Nevertheless, it is important to have a holistic idea about the common problems of those graphical policy authoring and visualization interfaces as a whole to develop more effective and usable policy authoring tools. Therefore, we conducted this SLR by considering all such interfaces we identified in the extracted literature to highlight their limitations.

However, even if graphical policy authoring interfaces guide the administrator to write correct and complete policies, those tools still fail to provide a complete solution to access control failures. Because failures can still occur due to human errors as the policy configuration using those tools is a manual and repetitive process \cite{narouei2017identification, kaur2021human, botta2007towards}. As a solution, previous research also suggested removing the human factor (i.e., the administrator) entirely from policy configuration by utilizing automated policy generation frameworks consisting of ML/NLP techniques \cite{shi2011controlled, xia2022automated}. However, those ML/NLP techniques employed in automated policy generation frameworks are not reliable enough to generate machine-executable policies without human supervision \cite{kaur2021human}. For example, ML models are often prone to false positives and false negatives. Therefore, if a generated policy with falsely identified policy components is applied to the access control system without being verified by an administrator, it causes security holes in the system. Those holes can open up a back door to hackers, resulting in data breaches. These kinds of limitations were also pointed out by other studies \cite{ del2022systematic}. As Del et al. revealed, neural network-based policy analysis techniques sometimes fail to identify privacy policies (i.e., low recall) \cite{del2022systematic}. At the same time, the traditional ML-based approaches tend to identify privacy policies falsely \cite{del2022systematic} (i.e., low precision). These findings support our claim that even if the automated solutions alleviate the administrator's overhead, they still are not reliable enough to operate without human involvement \cite{ del2022systematic, xia2022automated, heaps2021access, kaur2021human}. 

\begin{table}[]
    \centering
    \caption{Comparison of our SLR to other related survey/SLR articles. The glyph \faCircle  indicates the primary focus in the survey, \faAdjust indicates a secondary focus in the survey, and \faCircle[regular] indicates the respective aspect is briefly mentioned in the survey.}
    \label{tab: related}
    \begin{tabular}{ 
   >{\arraybackslash}m{6cm} 
   >{\centering\arraybackslash}m{1.5cm}
   >{\centering\arraybackslash}m{2.5cm}
   >{\centering\arraybackslash}m{2.0cm}}
        \toprule
         &  \textbf{Ours} & \textbf{Delaet et al.} \cite{delaet2010survey} &  \textbf{Paci et al.} \cite{paci2018survey} \\
        \midrule
          % \noalign{\vskip 2pt}
          \rowcolor{lightgray} \textbf{Considered aspect} &  &  &   \\
          \noalign{\vskip 2pt}
          Graphical access control policy configuration & \faCircle & \faCircle[regular] & \faAdjust   \\
          % \noalign{\vskip 2pt}
          Automated access control policy generation & \faCircle &  & \\
          % \rowcolor{lightgray} \textbf{Focus} &  &  &  & \\
          % Usability & \checkmark & \checkmark & \checkmark & \\
          % Reliability & \checkmark &  &  & \checkmark \\
          \rowcolor{lightgray} \textbf{Perspective} &  &  &   \\
          \noalign{\vskip 2pt}
          System administrator & \faCircle & \faCircle &    \\
          End user &  &  & \faCircle  \\

          \bottomrule
    \end{tabular}
\end{table}

Despite having those challenges in both the graphical policy authoring and visualization tools and automated policy generation frameworks, all the existing related surveys/systematic literature reviews have only focused on improving the usability of graphical access control policy configuration tools\cite{paci2018survey, delaet2010survey}. 
However, improving graphical policy configuration tools \cite{stepien2009non,stepien2014non, bertard2020using, turner2017proposed, brostoff2005r, zurko1999user, johnson2010usable, johnson2010optimizing, maxion2005improving, cao2006intentional, reeder2007usability, reeder2008expandable, morisset2018visabac, reeder2011more, karat2006evaluating, morisset2018building} alone will not mitigate access control failures as it induces accidental human errors due its manual and repetitive nature \cite{kaur2021human, narouei2017identification, narouei2017towards, nobi2022machine}. At the same time, improving fully automated policy generation frameworks \cite{shi2011controlled, alohaly2018deep, alohaly2021hybrid, fatema2016semi, slankas2013access, heaps2021access, slankas2013accessid, brodie2006empirical, xia2022automated, alohaly2019automated, xiao2012automated, narouei2018automatic, narouei2015automatic, slankas2012classifying, singru2020efficient, vaniea2008evaluating, rosa2020parser, inglesant2008expressions, liu2017automated, narouei2017identification, ye2016identify, perry2013modular, basile2010ontology, yang2021purext, slankas2014relation, narouei2017towards, alohaly2019towards, narouei2015towards, tanoli2018towards, zhu2021unstructured, brodie2005usable, shen2021using, abdelgawad2023synthesizing} alone will also not effectively reduce access control failure, as they are not 100\% accurate to operate without human supervision \cite{kaur2021human}. Therefore, while we agree the fact that both of those approaches should be improved by addressing their existing limitations, we further argue that the correct balance between manual and automated policy generation techniques will help develop more usable as well as reliable policy generation tools. To do that, first, the limitations of graphical policy configuration tools (i.e., manual approach) and the limitations of automated policy generation frameworks (i.e., automated approach) should be identified.

Nevertheless, to date, there is no publication that primarily analyzes the literature on both of the above approaches from the administrator's perspective and discusses their limitations, as shown in Table \ref{tab: related}. For example, according to Table \ref{tab: related}, even though Deleat et al. have focused on the administrator's perspective, they only briefly mentioned access control policy configuration tools in their survey \cite{delaet2010survey}. Furthermore, even though Paci et al. have analyzed the graphical access control policy configuration as a secondary focus in their SLR, they analyzed their usability from the end-user's perspective \cite{paci2018survey}. Therefore, to remedy this lack, we conducted this SLR by focusing on both the graphical policy authoring and visualization tools and NLP-based policy generation frameworks from the administrator's perspective, to point out their limitations. Addressing those limitations will help improve those tools and frameworks further and develop more effective policy generation frameworks that leverage the automation capabilities of reliable NLP techniques as well as the administrator's expertise via a usable interface. 

\section{Planning and Conducting the SLR}
\label{sec:slr}

To systematically identify the previous attempts in developing tools/frameworks to configure and generate access control policies and their limitations, we conducted a Systematic Literature Review (SLR) via a scientific and reproducible approach according to two main stages: planning the review and conducting the review \cite{kitchenham2004procedures}. 
% focusing on finding existing tools and frameworks developed for access control policy generation. 

\subsection{Planning the review}
\label{subsec:planning}

Planning the SLR is mainly about identifying the need for the review and developing a review protocol to avoid research bias \cite{kitchenham2004procedures}. With that in mind, we conducted the following activities under this section. 

\begin{enumerate}
    \item Defining the research scope
    \item Formulation of the research questions that need to be answered through the SLR
    \item Development of the search strings
    \item Selection of data sources 
    \item Definition of study selection criteria
\end{enumerate}

\subsubsection{Defining the research scope}

Defining the research scope is one of the crucial steps in planning the SLR \cite{papaioannou2016systematic}. It helps to formulate research questions, generate search strings, and establish the boundaries of the research \cite{del2018systematic}. Therefore, in order to determine the scope of this SLR, we adopted the PICOC (Population, Intervention, Comparison, Outcome, and Context) framework \cite{papaioannou2016systematic} as shown in Table \ref{tab: picoc}. Once the scope was defined, it was used to determine the approach to following SLR phases.

\begin{table*}{}
  \caption{SLR scope derived by following the PICOC framework \cite{papaioannou2016systematic, del2018systematic}.}
  \label{tab: picoc}
  \begin{tabular}{ 
   >{\centering\arraybackslash}m{1.8cm}
   >{\raggedright\arraybackslash}m{12.5cm}}
    \toprule
    \textbf{Concept} & \textbf{SLR application}\\
    \midrule
    Population & \underline{Access control}/\underline{privacy policy} \underline{configuration} and \underline{generation} from \underline{natural language} requirement specifications by the \underline{system administrator}. \\
    \noalign{\vskip 2pt}
    
    Intervention & Graphical policy authoring and visualization tools and NLP-based automated policy generation frameworks. \\
    \noalign{\vskip 2pt}
    
    Comparison(s) & Contrast between the graphical tools and NLP-based frameworks. \\
    \noalign{\vskip 2pt}
    
    Outcome & Existing interventions will be assessed by their usability, reliability, and contribution to the administrator's overhead and suggest guidelines to improve the above measures.  \\
    \noalign{\vskip 2pt}
    
    Context & Help administrators avoid access control failures  by improving the reliability and usability of existing policy configuration and generation tools. \\
    
    \bottomrule
  \end{tabular}
\end{table*}

\subsubsection{Formulation of research questions}
\label{subsubsec:rqs}

According to the scope we identified in Table \ref{tab: picoc}, the objective of this SLR is to investigate the prior work related to Access control/privacy policy configuration and generation from natural language requirement specifications by the system administrator. To achieve that goal, we developed the following research questions (RQs).

\textbf{RQ1}: \rqone

\textbf{RQ2}: \rqtwo

As the starting point, RQ1 focuses on investigating the existing tools and frameworks that are supposed to help the system administrator configure and generate access control policies from high-level access requirements often written in natural language \cite{narouei2017identification}. Then, in RQ2, we aim to ascertain the limitations of those tools and frameworks in order to develop effective frameworks to generate access control policies in the future. 

\subsubsection{Development of the search strings}
\label{subsubsec: searchstrings}

We identified a search string that is more likely to be used to extract the literature relevant to the area of interest. We used two techniques to decide which keywords to be included in each search string. 
\begin{enumerate}
    \item Extract keywords from the literature used for an initial assessment in the area of access control/privacy policy generation.
    \item Decide keywords based on the population in the defined scope as underlined in Table \ref{tab: picoc}.
\end{enumerate}
Based on the above two techniques, we decided our main keywords are "administrator", "access control", "privacy policy", "natural language", "generation", and "configuration". Even though we mainly focused on the access control domain, we considered "privacy policy" as one of the main keywords. Because in our initial literature assessment, we found that potentially relevant publications can also be found in the privacy policy domain. Apart from the above main keywords, we added several other keywords, such as "tool" and "interface", as part of our objective is to find tools or interfaces that are designed to help administrators generate or configure access control policies, even without knowing about complex access control languages or their syntax.

We used the wild card notation ("*") to include the different forms of the same word in the search process. For example, instead of using "administrator" we used "admin*" to represent different forms of the word, such as "administrators", "admin", "administration" and "admins".

According to the aforementioned criteria, the developed search string is as follows.

\noindent \texttt{admin* AND ("access control" OR "privacy policy" OR "privacy policies") AND ("natural language" OR ("natural language" AND "generat*") OR ("configur*" AND ("interface" OR "tool")))}

\subsubsection{Selection of data sources}
\label{subsubsec: datasources}

Publications were extracted from two main sources: 1) Scientific digital libraries and 2) Top-tier selected cybersecurity conferences. As the scientific digital library we mainly considered ACM Digital Library\footnote{https://dl.acm.org}, IEEE Xplore Digital Library\footnote{https://ieeexplore.ieee.org/Xplore/home.jsp}, Springer Link\footnote{https://link.springer.com}, and Elsevier Science Direct\footnote{https://www.sciencedirect.com}. Google Scholar\footnote{https://scholar.google.com} was used only to extract publications in Phase 3 of Section \ref{subsubsec: studySelect}. We found that the way we should define the search string for each search engine is different. For instance, ACM provides means to enter keywords in the search string separately and filter the literature based on the venue, even though digital libraries such as Elsevier Science Direct only allow the user to enter the complete search string in the search box. 
Furthermore, as we later found out, Elsevier Science Direct digital library will not accept search strings that contain wild card notation and strings containing more than eight boolean operators. Therefore, we used, 

\noindent \texttt{"administrator" AND "access control" AND ("natural language" OR ("natural language" AND "generation") OR ("configuration" AND "tool"))}

as the search string in that occasion to search publications.  

To search for exact matches for the search string keywords, quotation marks ("") were used (e.g., "access control"). In addition, to ensure the keywords are searched together, we combined the keywords with boolean operators such as AND. Even though we only used the AND and OR operators, the other basic operator (NOT) is also available in the above platforms to be used. 

Apart from the literature we extracted from the scientific digital libraries, we searched through the top-tier selected cybersecurity, human-computer interaction (HCI), as well as natural language processing conferences and journals relevant to the topics in this SLR. Even though this SLR is not directly related to Human-Computer Interaction (HCI), we considered such conferences because the policy generation component involves natural language processing techniques to translate human intentions to machine (computer) executable policy. We selected eight conferences and journals, namely, IEEE Symposium on Security and Privacy (IEEE S\&P), USENIX Security, ACM Conference on Computer and Communications Security (CCS), ACM Transactions On Privacy and Security (ACM TOPS), Transaction of the Association for Computational Linguistics (TACL), Network and Distributed System Security Symposium (NDSS), ACM Conference on Human Factors in Computing Systems (CHI), Symposium on Usable Privacy and Security (SOUPS) and Conference on Empirical Methods in Natural Language Processing (EMNLP). The breakdown of the selected venues is shown in Table \ref{tab: venues}.

\begin{table*}[]
  \caption{Selected venues.}
  \label{tab: venues}
  \begin{tabular}{ 
   >{\centering\arraybackslash}m{2cm} 
   >{\centering\arraybackslash}m{6cm} 
   >{\centering\arraybackslash}m{2cm}
   >{\centering\arraybackslash}m{2cm}}
    \toprule
    \textbf{Type} &  \textbf{Cybersecurity} & \textbf{HCI} & \textbf{NLP}  \\
    \midrule
    Conferences & IEEE S\&P, CCS, NDSS, SOUPS & CHI & EMNLP\\
         Journals & USENIX, ACM TOPS & - & TACL\\
    
    \bottomrule
  \end{tabular}
\end{table*}

\subsubsection{Definition of study selection criteria}
The study selection criteria help to determine what studies should be included and excluded from the SLR \cite{kitchenham2004procedures, fink2019conducting}. Therefore, we defined the inclusion (IN) and exclusion (EX) criteria as shown in Table \ref{tab: criteria} to be relevant according to the scope in Table \ref{tab: picoc}.

\begin{table*}[]
  \caption{Inclusion and Exclusion Criteria}
  \label{tab: criteria}
  \begin{tabular}{ 
   >{\centering\arraybackslash}m{0.5cm} 
   >{\raggedright\arraybackslash}m{12.8cm}}
    \toprule
    \textbf{ID} &  \textbf{Criteria}\\
    \midrule
    IN1 & The publication presents either an automated access control policy generation framework from natural language or a tool to write and/or visualize access control policies.\\

         % IN2 & The publication presents a top-down policy engineering framework.\\

         IN2 & The publication should be related to access control or privacy/security policy authoring and visualization done by the system administrator\\

         IN3 & The publication has been peer-reviewed.\\ 

         IN4 & The publication was written in English.\\

         \midrule
         \noalign{\vskip 4.5pt}
         EX1 & The publication is a secondary study.\\

         EX2 & The publication presents a concept that has not been implemented and tested yet. \\

         EX3 & The publication only presents a NLP technique for information extraction without applying it to access control or privacy policy domain.\\

         EX4 & The publication mainly focuses on an access control model or a language but not on a policy configuration tool, or an automation framework.\\

         EX5 & The publication focuses on avoiding access control failures either from the end user's perspective or the software developer's perspective. \\

         EX6 & The publication presents a bottom-up policy mining/text mining approach.\\

         EX7 & The publication focuses on information extraction from privacy notices.\\
         
         EX8 & The publication presents a tool designed to write policies in a standard policy language by following their same strict syntax (e.g., XML editors to write XACML policies \cite{nergaard2015scratch}).\\
    
    \bottomrule
  \end{tabular}
\end{table*}

\subsection{Conducting the review}
\label{subsec:conductingSLR}

\begin{figure*}[]
    \centering
    \includegraphics[height=9cm]{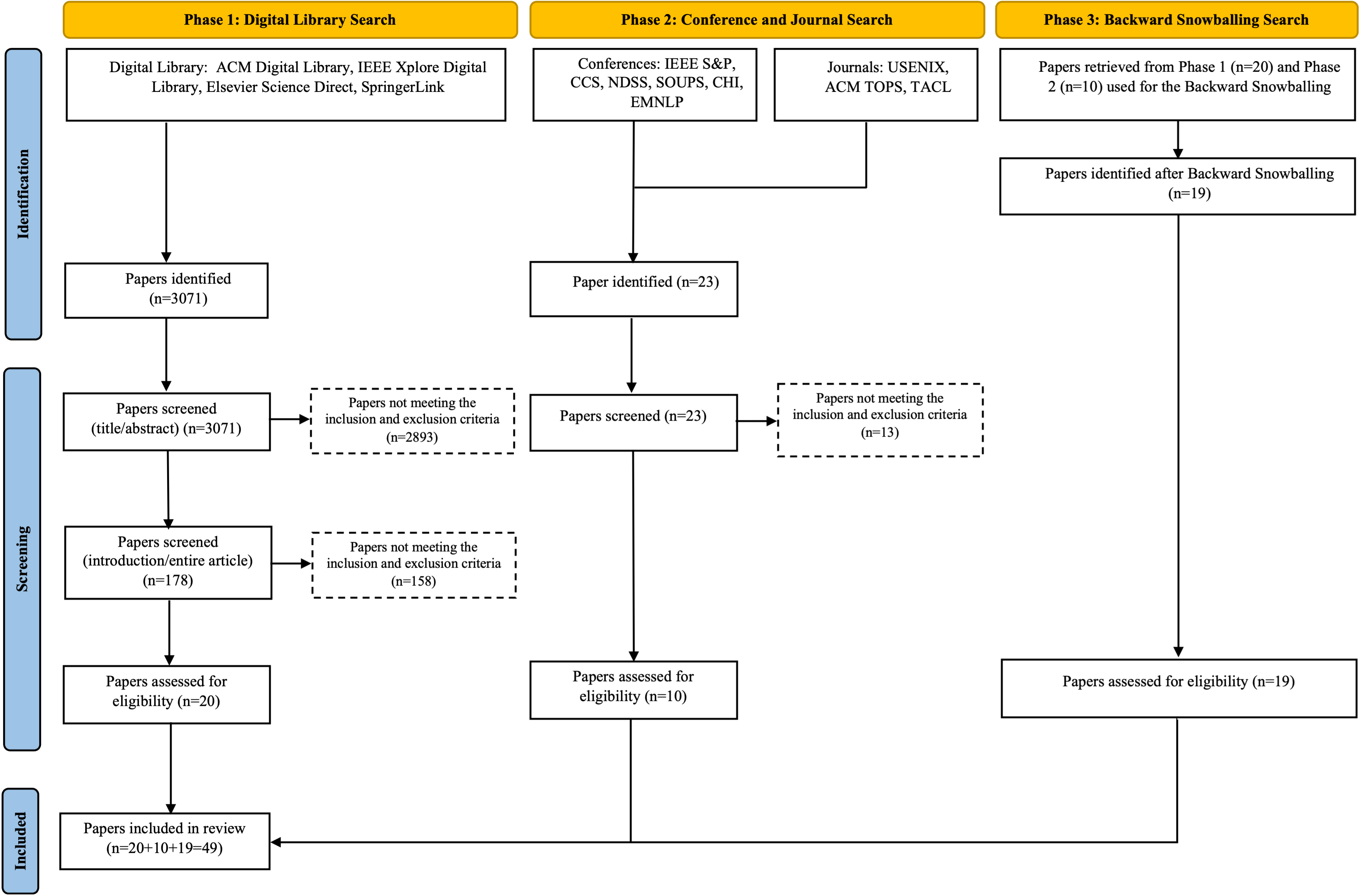}
    \Description[PRISMA flow diagram showing the research identification and study selection process.]{The image shows a PRISMA flow diagram that summarizes how the particular studies were identified and selected with the number of papers in each step. The retained numbers of papers are shown inside solid-lined boxes, and the numbers of papers we excluded are displayed within dash-lined boxes. The identification was carried out in 3 phases: digital library search, conference and journal search, and backward snowballing, as mentioned at the top of the figure. In the digital library search, we first identified 3071 papers using the search string developed. After screening the title and abstract based on the inclusion and exclusion criteria, we left with 178 papers to screen further. Then, based on the inclusion and exclusion criteria, we excluded another 158 papers by retaining 20 papers to read. In the conference and journal search, we identified 23 papers that were not included in the previous phase. After filtering those using the inclusion and exclusion criteria, we were left with another ten papers to read. As the last step, we used the 20 papers we kept from phase 1 and 10 papers returned from the previous phase to conduct backward snowballing. Backward snowballing returned 19 papers, bringing the final paper count up to \pc as the included publications in this SLR.}
    \caption{PRISMA \cite{page2021prisma} Flow diagram summarizing the research identification and study selection process.}
    
    \label{fig:prisma}
\end{figure*}

After planning, SLR is conducted. As suggested by \cite{kitchenham2004procedures}, we performed three main activities, namely \emph{research identification and study selection, data extraction, and data analysis}. 

\subsubsection{Research identification and Study Selection}
\label{subsubsec: studySelect}

This activity was performed from November 2022 - August 2023 under three main phases according to the PRISMA framework \cite{page2021prisma} as shown in Fig. \ref{fig:prisma}. 
\begin{itemize}
    \item \textbf{Phase 1: Digital Library Search - }We searched each library mentioned in the Section \ref{subsubsec: datasources} by applying the search strings developed under the Section \ref{subsubsec: searchstrings}.

    \item \textbf{Phase 2: Conference and Journal Search - }In order to avoid the publication bias \cite{kitchenham2004procedures}, we scanned through the selected journals and conferences mentioned in Section \ref{subsubsec: datasources} by applying the search strings developed under the Section \ref{subsubsec: searchstrings}.

    \item \textbf{Phase 3: Backward Snowballing Search - }To ensure the relevant publications are not overlooked, we searched through the references and citations in the retrieved publication from Phase 1 and Phase 2~\cite{wohlin2014guidelines}. 
\end{itemize}

In Phase 1 (Search across the Digital Libraries), the developed search string returned a total of 3071 publications from 2013 to 2023. Even though we limited our search from 2013 to 2023, it does not mean we did not consider relevant publications before 2013. Phase 3 allowed us to retrieve such publications using the backward snowballing technique. We screened the returned publications in 2 steps. First, after reading the title and abstract of the identified publications, we added 178 publications that match the inclusion and exclusion criteria to Zotero Reference Management Software\footnote{https://www.zotero.org} by removing 2893 articles from consideration at the beginning.
% After removing duplicates using Zotero Reference Management Software\footnote{https://www.zotero.org}, a dataset of 1290 publications was obtained. 
% Then, 1290 extracted publications 
Secondly, those retained 178 publications were further reviewed by reading the abstract, introduction, and sometimes the entire publication to decide whether or not to include it in the SLR. After removing the publications that did not match the inclusion and exclusion criteria, 20 publications were left in the end as a result of Phase 1. 

Searching through the selected conferences and journals mentioned in Table \ref{tab: venues} was done under Phase 2. Our search was limited to publications from 2013. By using the same search strings, we extracted 23 potentially relevant publications. It is worth mentioning that we were careful not to extract publications that were already extracted in Phase 1. Once the publications were reviewed, we obtained ten publications after applying the inclusion and exclusion criteria. 

Additional papers were identified by Backward Snowballing Search \cite{wohlin2014guidelines} in Phase 3. This method allows us to go through the references and citations mentioned in the publications obtained in Phase 1 and Phase 2. We extracted 19 relevant papers in this phase without any time limitations. At the end of Phase 3, we have a total of \pc publications retrieved by SLR. The summary of the research identification and study selection process is depicted in Fig. \ref{fig:prisma}, as suggested by \cite{desolda2021human}.

\subsubsection{Data Extraction}

% The 53 publications which were obtained using the 3 phases mentioned in Section \ref{subsubsec: studySelect}, are listed in Table \ref{tab:pubs} sorted according to the surname of the first author. 

% There are several observations made while selecting and extracting the publications. Removing the search term "admin*" from the search string significantly increased the number of publications returned. Therefore we were able to extract publications that can potentially be helpful for the access control policy generation, even though they do not directly address the system administrator. 
When extracting the publications, as we included the terms "privacy policies" and "privacy policy" into the search string, multiple publications that extract information from "privacy notices", such as data practices, were returned. In almost all of those publications, the authors' main focus was to develop techniques to extract information from lengthy privacy notices and present them to the \textbf{end user} in a more usable way. Therefore, since the returned publications were neither related to the access control policy configuration domain nor focused on the system administrator, we decided to exclude them. However, privacy notices also contain natural language access control policies that can be used to extract information for policy generation \cite{brodie2006empirical, brodie2005usable}. Therefore, after careful consideration, we retained several publications related to the "privacy policy" domain that focus on extracting access control policies and their rules. 

In order to extract the necessary data from the included publications, we first examined them. From each publication, we collected general quantitative data (e.g., title, author, year, published venue, etc.) as well as the qualitative information that aligns with the formulated research questions in Section \ref{subsec:planning}. 
% Even though, he reviewed the entire publication, he focused on several particular sections in the publication to extract necessary information to answer the research questions. 

% \begin{itemize}
%     \item \emph{RQ1}: Abstract, Introduction, Proposed methodology, Conclusion.
%     \item \emph{RQ2}: Threats to validity, Challenges, Experimental Results, Evaluation, Discussion, Related Works, Conclusion.
    
% \end{itemize}

\subsection{Data Analysis}
\label{subsubsec: data_analysis}

Since we aim to answer the research questions qualitatively by identifying recurring patterns of meaning and concepts within the extracted data, next, we conducted \textbf{Thematic Analysis} \cite{braun2012thematic} according to the following steps. 

\begin{enumerate}
    \item \emph{Familiarisation}: Extracted data was read and summarised using the Mendeley Reference Management Software\footnote{https://www.mendeley.com} to have an overview of the data. We used Mendeley instead of Zotero for this step, as it offers more advanced annotation and note-taking features.
    
    \item \emph{Coding}: As we are coding with specific research questions (i.e., RQ1 and RQ2) in mind, we followed the integrated coding procedure \cite{cruzes2011recommended}, which allows us to start coding with an initial list of codes, derived based on research questions and authors' knowledge on access control policy generation (deductive) and expand the code list by adding new codes based on the extracted data (inductive). By following that approach, the first author assigned codes that reflect the relevant features to answer the research questions. 
    % However, we did not use line-by-line coding technique as the first author was supposed to code more than 50 publications. Instead of line-by-line coding he relied on extracted data and the written summary to assign codes. 
    For instance, the first author assigned the code \emph{"parsing"} to the text \emph{"In our approach, after identifying the different sentence types, we parse each line (sentence) using the Stanford Natural Language Parser ..."} \cite{slankas2014relation}. Furthermore, the extracted data were read multiple times to refine the codes and ensure they were assigned correctly.
    
    \item \emph{Generating initial themes}: Upon coding, all the codes were compiled into logical groups (e.g., text editors, templates, etc.) to identify themes that help to answer the research questions (RQ1 and RQ2). 
    
    \item \emph{Reviewing the themes}: Initial themes were checked against the data segments extracted to ensure that they tell a compelling story about policy generation tools, frameworks, and their limitation with the involvement of all the authors. To fine-tune the story, we refined the initial themes and sometimes split existing themes. For example, we split the theme graphical policy configuration into policy authoring and policy visualization to highlight the different approaches that help administrators write and understand policies.
    
    \item \emph{Defining and naming higher order themes}: Finally, we defined two main themes: graphical policy authoring and visualization and NLP-based automated policy generation, with their detailed descriptions, and assigned each initial theme to one of those two categories.
    
    % The relationships between the initial themes were used to define overarching higher order themes with a meaningful name and a detailed description. 
    % For instance, the themes for RQ1 were generated based on the policy authoring guidance and automation solutions. (e.g. GUI based policy authoring, Natural language based policy translation)
\end{enumerate}

The co-authors validated the process by reviewing the consistency of codes and themes against the associated data and examining whether the generated themes respond to the research questions RQ1 and RQ2. Several meetings were conducted involving the three authors to discuss the disagreements and issues of generated codes and themes. As a result, we minimized the possible inconsistencies in the coding process. Once the themes and categories were generated, the main author filled out a spreadsheet to classify the articles based on the detailed descriptions of the themes/categories. Later, the agreements and disagreements of the other coders regarding the classification decisions were expressed and recorded in a meeting to calculate the metrics for inter-rater reliability (i.e., Cohen kappa ($\kappa$) \cite{viera2005understanding} and percentage of agreement). After the calculation, we noticed that the co-authors show a "substantial agreement" ($\kappa$ $\geq$ 0.76) in classifying articles into the defined categories. However, we came across several disagreements on classifying based on the graphical policy configuration tools (authoring vs. visualization tools). We discussed the disagreements in a meeting and resolved them by examining the descriptions of the categories. Even though we measured the inter-rater reliability here, it is worth noting that we were more focused on incorporating different perspectives of co-authors when developing themes and assigning relevant publications to them than the reliability measurement, as advised by Braun and Clark \cite{braun2012thematic}.

\section{Results}
\label{sec: results}

\begin{figure*}[]
    \centering
    \includegraphics[height=17.5cm]
    {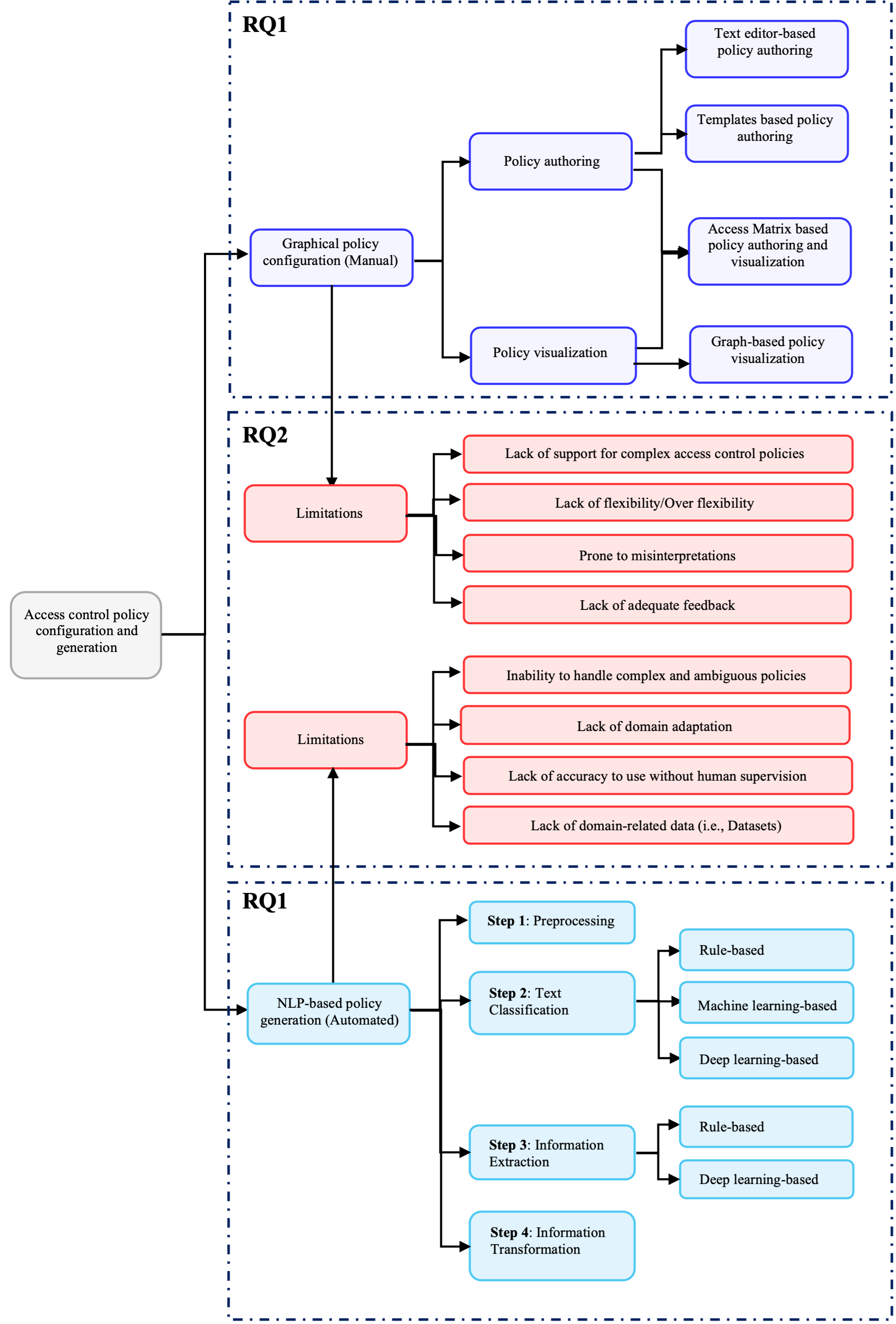}
    
    \caption{Thematic map showing the identified access control policy configuration and generation approaches. Under RQ1, we discuss graphical policy configuration (i.e., manual) in terms of policy authoring and visualization tools and NLP-based policy generation (i.e., automated) in terms of 4 steps. Under RQ2, we discuss the limitations of identified tools and frameworks related to the above approaches.}
    \Description[Thematic map showing how the access control policy configuration was done previously and their limitations]{The image shows a tree summarizing the existing access control policy configuration and generation techniques and their limitations. Access control configuration can be done in terms of graphical policy authoring and visualization and NLP-based automated policy generation. Graphical policy authoring and visualization are further divided into two techniques: policy authoring and policy visualization. Policy authoring can be done using text-editor-based and template-based policy authoring tools, while policy visualization can be done using access matrix-based tools and graph-based visualization tools. Access matrix-based tools belong to both policy authoring and visualization techniques. Then, the limitations of the aforementioned tools are shown in the image inside a dashed box named RQ2. They are the lack of support for complex access control policies, the lack of flexibility, and inconsistent interpretations. On the other hand, NLP-based automated policy generation frameworks can be done in 4 steps, namely, pre-processing, text classification, information extraction, and information transformation, which are displayed as leaf nodes of the tree. Similar to the graphical policy authoring and visualization tools, limitations of automated policy generation frameworks are also displayed in the same dashed box named RQ2 as before. They are the inability to handle ambiguities, the lack of flexibility, the lack of accuracy, and the lack of domain-related data.}
    \label{fig:tools}
\end{figure*}

Access control failures due to policy configuration mistakes by system administrators can result in drastic data breaches, putting an entire organization at risk of financial loss and reputation damage \cite{bauer2009real}. As we identified, such access control failures occur mainly due to the lack of usability and lack of reliability of the existing access control policy configuration and generation solutions \cite{stepien2009non, johnson2010optimizing, reeder2008expandable, morisset2018building, slankas2014relation, xia2022automated, xiao2012automated}. Therefore, to avoid such failures, it is essential to improve their usability and security in terms of reliability (i.e., address usability-security trade-off). To achieve that, first, we have to identify where those usability and reliability issues are in the existing tools and frameworks and provide solutions for them. With that motivation, we conducted this SLR identifying two main types of tools and frameworks developed to configure and generate access control policies
% even without having the knowledge of access control models, languages, and syntax
: \textbf{Graphical policy configuration (i.e., authoring and visualization) tools} \cite{stepien2009non,stepien2014non, bertard2020using, turner2017proposed, brostoff2005r, zurko1999user, johnson2010usable, johnson2010optimizing, maxion2005improving, cao2006intentional, reeder2007usability, reeder2008expandable, morisset2018visabac, reeder2011more, karat2006evaluating, morisset2018building} (e.g., text editor-based tools, template-based tools, access matrix-based tools, and graph-based tools), \textbf{NLP-based automated policy generation frameworks} \cite{shi2011controlled, alohaly2018deep, alohaly2021hybrid, fatema2016semi, slankas2013access, heaps2021access, slankas2013accessid, brodie2006empirical, xia2022automated, alohaly2019automated, xiao2012automated, narouei2018automatic, narouei2015automatic, slankas2012classifying, singru2020efficient, vaniea2008evaluating, rosa2020parser, inglesant2008expressions, liu2017automated, narouei2017identification, ye2016identify, perry2013modular, basile2010ontology, yang2021purext, slankas2014relation, narouei2017towards, alohaly2019towards, narouei2015towards, tanoli2018towards, zhu2021unstructured, brodie2005usable, shen2021using, abdelgawad2023synthesizing} that often contain four steps (e.g., (1) pre-processing, (2) text classification, (3) information extraction, and (4) information transformation), and their limitations, as shown in Fig. \ref{fig:tools}. Identifying those limitations will help discover ways to improve the usability and reliability of access control policy generation and, in turn, help develop more effective access control policy generation frameworks in the future.

% A summary of the results can be found at \url{https://cutt.ly/iwkj3Z8x}.

\subsection{Graphical policy authoring and visualization tools}
\label{subsubsec: gui}

According to our thematic analysis \cite{braun2012thematic}, 16 of the included publications proposed graphical user interfaces (GUIs) to guide administrators in access control policy authoring and visualization. Among those publications, some have proposed \textbf{text editors} \cite{stepien2009non, stepien2014non, brostoff2005r, zurko1999user, shi2011controlled, inglesant2008expressions} and \textbf{template-based authoring tools} \cite{johnson2010optimizing, johnson2010usable, rosa2020parser, turner2017proposed, brodie2005usable, brodie2006empirical, karat2006evaluating, vaniea2008evaluating} to help administrators to write access control policies without even having knowledge about policy languages such as XACML (Extensible Access Control Markup Language) \cite{stepien2009non}. On the other hand, some have also proposed \textbf{graphs-based policy visualization} \cite{morisset2018building, morisset2018visabac, bertard2020using} and \textbf{access matrices-based policy authoring and visualization tools} \cite{reeder2007usability, reeder2008expandable, maxion2005improving, cao2006intentional} that allow the administrator to visualize the policies with a low cognitive load \cite{vaniea2008access}. 
% Therefore in this section, we discuss the above types of graphical policy authoring and visualization tools and their unique features as we identified in the SLR.

\subsubsection{\textbf{Text editor-based policy authoring tools}}
\begin{figure*}[]
    \centering
    \includegraphics[height=9cm]{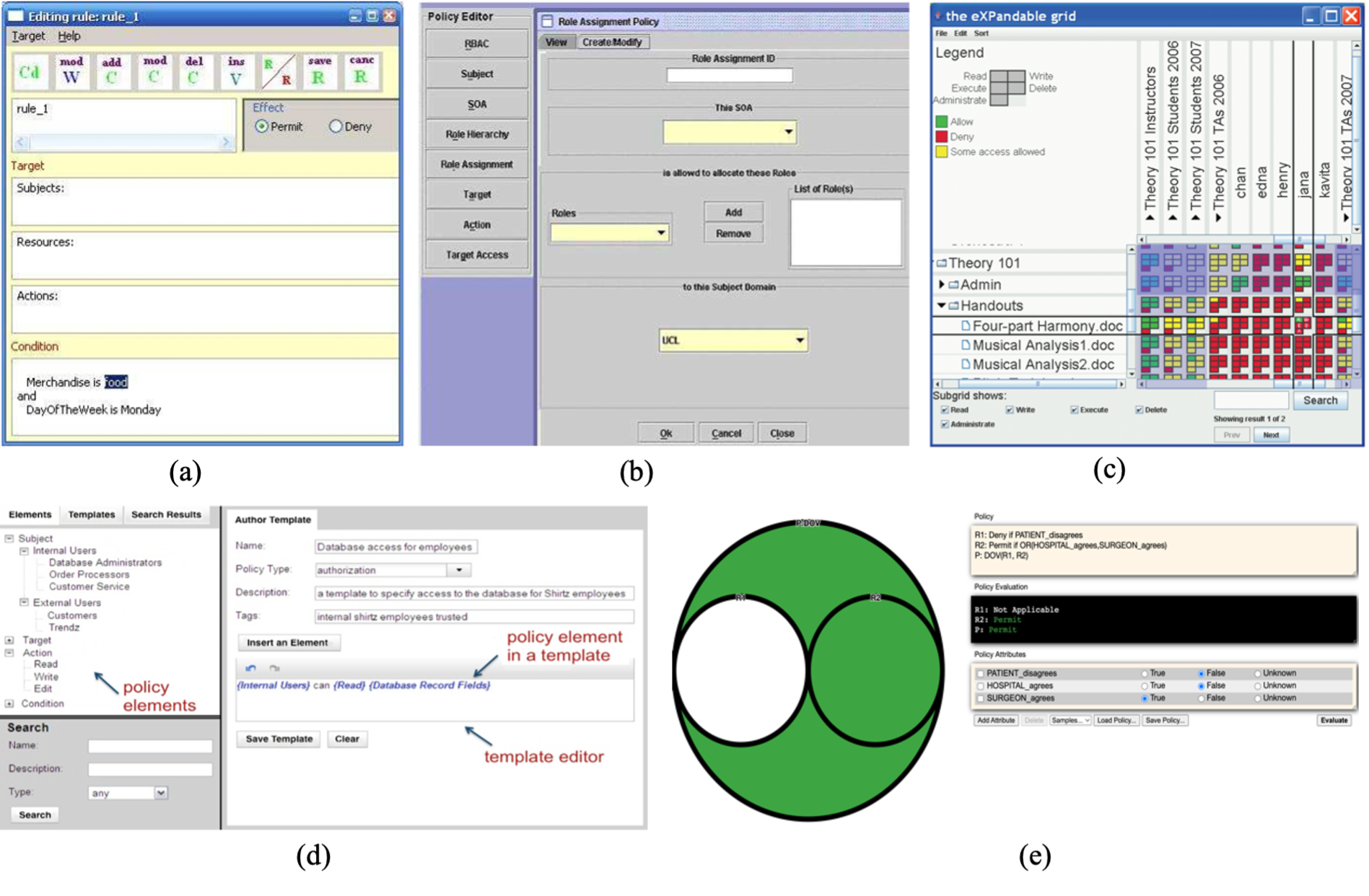}
    \caption{Examples for graphical policy authoring and visualization tools: (a) "easyXACML" text-editor-based tool \cite{stepien2009non}, (b) Policy authoring interface developed for e-scientists \cite{brostoff2005r} (c) "Expandable Grids" access matrix-based visualization tool \cite{reeder2008expandable, reeder2011more}, (d) Template-based tool by Johnson \cite{johnson2010usable}, (e) "VisABAC" visualization tool \cite{morisset2018building}.}
    \Description[Examples for Graphical User Interfaces]{The image contains five graphical policy authoring and visualization interfaces as sub-figures. The first sub-figure (a) shows a text editor-based graphical policy authoring interface that contains five text boxes to write the rule id, subject, action, resource, and conditions in controlled natural language. It also contains a radio button to select whether a written policy is an allow or a deny policy. The second sub-figure (b) also contains a text editor-based interface that allows administrators to choose correct policy components from the LDAP directory structure. The third sub-figure contains an access matrix-based interface containing a matrix that denotes resources in columns and users/groups in rows. The fourth sub-figure (d) contains a template-based policy authoring interface that provides policy templates to write policy components by adding policy components to the placeholders. The last sub-figure contains a treemap-based policy visualization interface. In that sub-figure left side contains the circular tree map describing the rules and the policy shown on the right side of the image.}
    \label{fig:gui_tools}
    
\end{figure*}

% \textbf{Text editor-based policy authoring tools} - 
Among the identified graphical policy authoring and visualization tools, four of them are text editor-based policy authoring tools \cite{stepien2009non, stepien2014non, brostoff2005r, zurko1999user}. Instead of letting the administrator write the access control policy directly in a policy language, policy editors allow them to build the policy just by defining the values for respective policy components (i.e., subject, actions, resource, etc.) either using separate text boxes \cite{zurko1999user, stepien2009non, stepien2014non} as shown in Fig. \ref{fig:gui_tools}(a) or using drop-down menus \cite{brostoff2005r} as shown in Fig. \ref{fig:gui_tools}(b). 

However, the identified text editor-based policy authoring tools do not often support policies containing multiple rules with different access decisions. For instance, consider a hypothetical policy containing two rules, such as \emph{"The nurse cannot write the patient's records to prescribe medicine, but doctors can, only if the patient agrees."}. If the above policy is to be written using an identified text editor-based policy authoring interface \cite{stepien2009non, stepien2014non, brostoff2005r}, the administrator has to extract the two rules of the policy first. Then, write them separately using the policy authoring tool to build the policy, as one rule is an allow rule and the other is a deny rule. When writing individual rules manually to build a policy, it could result in mistakes in those rules, leading to access control failures. For example, since some text editor-based tools allow administrators to write policy components by themselves \cite{stepien2009non, stepien2014non}, sometimes they use different terms in each rule to represent the same component in the policy, such as "patient's records" in the rule associated with the nurse and "records" in the rule associated with the doctor \cite{reeder2007usability}. However, the resource "records" can have a broader scope compared to more specific "patient's records" \cite{brostoff2005r}. As a result, since the doctors were given inappropriate access, they can access any record and alter or steal it without alerting the administrators, causing data breaches \cite{bauer2009real, brostoff2005r, Page_2023}. This limitation is defined as the "policy component problem" by Brostoff et al., as they revealed novice administrators often fail to identify correct policy components and misinterpret their scope, leading them to write incorrect policies using text editor-based tools, creating security holes in the authorization system \cite{brostoff2005r}.

Another limitation of existing text editor-based policy authoring tools is their functionalities are often prone to misinterpretations as they do not provide all the relevant information to write a policy explicitly, such as default rules \cite{brostoff2005r, reeder2007usability}. For example, since Brostoff et al. did not clearly mention that their text editor-based interface operates under the default-deny principle, their user study participants wrote deny rules explicitly, producing redundant rules \cite{brostoff2005r}. They defined this problem as the "policy paradigm problem" \cite{brostoff2005r}.

Furthermore, the existing text editor-based tools \cite{stepien2009non, stepien2014non, brostoff2005r} are not flexible enough to support access control policies that could contain other types of policy components such as purposes and obligations \cite{reeder2007usability}. As a result, even if the policy should contain a purpose that has to be checked before granting/denying access, administrators might neglect it when building policies using text editor-based policy authoring tools. In that case, according to the aforementioned example, doctors will be able to edit the patient's records for any reason, causing access control failures.
% As a result, a third party might be able to exploit such security holes by using doctors' credentials to steal patients' private medical records, causing data breaches. 
Therefore, in order to avoid such scenarios, template-based policy authoring tools were developed by providing pre-defined templates that support policy components necessary for a particular organization to write policies using pre-defined policy elements \cite{johnson2010optimizing, johnson2010usable, turner2017proposed}.

\subsubsection{\textbf{Templates-based policy authoring tools}}
% \textbf{Templates-based authoring tools} - 
A "template" is a specification of the structure of a list of natural language policies \cite{johnson2010usable}. Template-based access control policy authoring tools provide one or more such templates with placeholders to input policy components to build a complete access control policy \cite{johnson2010optimizing, johnson2010usable, turner2017proposed}. For example, to write a policy such as \emph{"Database administrators can read database record fields."}, Johnson et al. \cite{johnson2010usable} has provided a template of \texttt{\{Internal Users\} can \{Action\} \{Resource\}} as shown in Fig. \ref{fig:gui_tools}(d).
However, a single template would not suffice to support all the policy requirements of an organization. Therefore, Johnson et al. also have allowed in their policy authoring interface shown in Fig. \ref{fig:gui_tools}(d), to create and modify new templates depending on the access requirements \cite{johnson2010optimizing, johnson2010usable}. 

Nevertheless, user study participants have raised concerns that the template-based policy authoring tool of Johnson et al. is overly flexible as it might allow administrators to make "general" templates \cite{johnson2010usable}. For example, if the template is \texttt{\{Subject\} can \{Read\} \{Database record fields\}}, the administrator can input any user (e.g., internal users and external users) as the \texttt{Subject} allowing even external users to read database records, even if it should only be done by internal users as shown in Fig. \ref{fig:gui_tools}(d). If external users are able to read databases containing the organization's confidential information, such as customers' personal details, it will result in data breaches, harming the organization and its customers \cite{Page_2023}. On the other hand, some other template-based policy authoring interfaces \cite{turner2017proposed} lack the flexibility to support the unique access requirements of different organizations, as they are limited by one pre-defined template specifically designed for one particular type of rules, such as business rules \cite{turner2017proposed}. Therefore, if the policy requirements are different in other domains, such as healthcare \cite{rosa2020parser}, the same template-based tool cannot be easily adapted to support those different access control policies. 

However, neither text editor-based tools nor template-based policy authoring tools help administrators to graphically visualize the existing policies so that administrators can easily understand the relationships between policies \cite{maxion2005improving, bertard2020using}. That understanding is important, especially when making changes to existing policies \cite{reeder2008expandable, maxion2005improving}. For instance, consider a scenario where the administrator has to add a user who is restricted from reading the financial information of an organization to a group that has permission to read that information. In that case, if the administrator cannot clearly see that the user is going to be added to a group with conflicting permissions and the allow rules take precedence \cite{reeder2011more}, the administrator might accidentally allow that user to access the financial information of the organization, leading to access control failures \cite{reeder2011more, reeder2008expandable}. 

Therefore, to avoid such failures due to the lack of holistic awareness of how the policies affect each other, previous literature also proposed access matrix-based policy authoring and visualization tools \cite{reeder2008expandable, reeder2011more, maxion2005improving, cao2006intentional} and graph-based policy visualization tools \cite{morisset2018building, morisset2018visabac, bertard2020using} to visualize access control policies.

\subsubsection{\textbf{Access Matrix-based policy authoring and visualization tools}}
% \textbf{Access Matrix-based policy authoring and visualization tools} - 
"Access control matrix" is the most common type of interface we identified that guides the administrator to both configure access control policies and visualize them in a matrix. Lampson first introduced it \cite{lampson1974protection} as a two-dimensional table, where each row represents a user, each column represents a resource, and each cell contains operations that the subject is allowed to perform on the resource. After that, the access control matrix-based tools were proposed as a method of configuring and visualizing access control policies \cite{reeder2008expandable, reeder2011more, cao2006intentional, maxion2005improving}. For example, Reeder et al. proposed "Expandable Grids" shown in Fig. \ref{fig:gui_tools}(c) \cite{reeder2008expandable, reeder2011more}. It represents all the users and their associated groups in columns and all the resources and associated resource groups in rows. In contrast to the other access matrix-based interfaces \cite{maxion2005improving, cao2006intentional}, Expandable Grids uses a sub-grid in each cell to authorize each user to read, write, delete, execute, and administrate resources. Furthermore, to represent whether the above five actions are allowed, denied, or some access allowed when a user inherits access from a group (i.e., effective permission), green, red, and yellow colors were used, respectively, as shown in Fig. \ref{fig:gui_tools}(c).

Nevertheless, since access matrices are two-dimensional, they can only represent users and resources by their two axes. Therefore, access matrix-based policy authoring and visualization tools often cannot represent access control policies that contain policy components such as conditions and purposes, even though they significantly affect the authorization decision \cite{yang2021purext, alohaly2019towards}. For instance, consider a similar example we used earlier \emph{"The doctor can read the patient's records to prescribe medicine only if the patient agrees."}. Suppose the condition (i.e., if the patient agrees) and the purpose (i.e., to prescribe medicine) are neglected because access matrices do not support them. In that case, the configured policy implies that the doctors neither need the patient's consent nor a reason to access and write the patient's medical records. As a result, not only does it cause a privacy violation, but also anyone with doctors' credentials can gain access to someone's medical history easily. 
Furthermore, the access matrix-based visualization approach can be cumbersome when dealing with a large number of users and resources in the organization, resulting in policy misconfigurations \cite{morisset2018visabac, morisset2018building}. For example, consider an access matrix containing many users (i.e., columns) and many resources (i.e., rows). To give permission to a user to access a resource, the administrator might need to navigate through many columns to find the correct user and follow the column through many rows using the mouse until the correct resource is found. When navigating, if the administrator's mouse accidentally slips from the desired row/column to an adjacent row/column, the administrator might incorrectly identify the wrong resource and/or wrong user and give permission, causing an access control misconfiguration (i.e., "off-by-one error" \cite{reeder2008expandable}).  
 Therefore, to avoid such scenarios due to difficulties of navigation of the conventional access matrix, graph-based policy visualization tools were developed \cite{morisset2018building, morisset2018visabac}.

 \begin{figure}[]
  \centering
    \includegraphics[height=2cm]{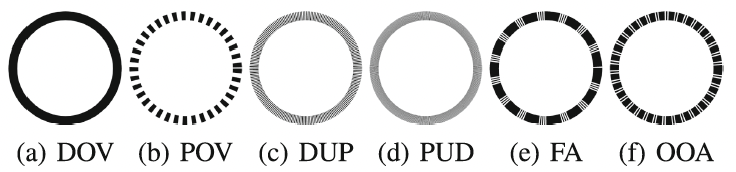}
    \caption{Line conventions used by VisABAC to represent operations (a) Deny overrides (b) Permit overrides (c) Deny unless permit (d) Permit unless deny (e) First applicable (f) Only one applicable \cite{morisset2018building, morisset2018visabac}. The line styles used to represent Permit Unless Deny and Deny Unless Permit look almost similar, increasing the chance to misidentify.}
    \Description[Line styles used in VisABAC interface]{The image shows line styles used in VisABAC interface to denote rule combine algorithms. However, it is hard to identify the line styles used to represent deny-unless-permit and permit-unless-deny scenarios separately, leading to misinterpretations in the policy visualization.}
    \label{fig: visabac}
\end{figure}

\subsubsection{\textbf{Graph-based policy visualization tools}}
% \textbf{Graph-based policy visualization tools} - 
Graphs/trees can be in different forms, such as layered graphs \cite{bertard2020using}, and treemaps \cite{morisset2018building}. Each type of graph has nodes and edges that connect nodes. In some policy visualization tools, nodes represent policy components such as users, and edges represent relationships between them \cite{bertard2020using}. In several other cases, nodes represent rules in a policy, and edges represent the relationships between those rules and how they are being combined to form the policy \cite{morisset2018building, morisset2018visabac}. For example, Morisset et al. introduced a treemap-based visualization tool, "VisABAC" to visualize Attribute-based Access Control (ABAC) policies \cite{morisset2018building, morisset2018visabac} as shown in Fig. \ref{fig:gui_tools}(e). They utilized a special form of treemaps named "Circular Treemap," which represents tree nodes as circles. Therefore, a parent node with two children nodes will be represented as a circle containing two sub-circles in the circular treemap. In the access control domain, the parent circle would be a policy, and the children circles would be either policies or rules that are being combined to form the policy \cite{morisset2018visabac, morisset2018building}. In contrast to access matrices, VisABAC focuses on rules, policies, and their relationships instead of policy components such as subjects, actions, and resources. Therefore, no matter how high the number of users and resources in the organization is, it does not affect the visualization, making the visualization easy to navigate compared to access matrices \cite{morisset2018building}.
% Moreover, to visualize RBAC policies, Bertard et al. proposed a layered graph-based approach named "Sugiyama-styled Graph" in \cite{bertard2020using}. The proposed approach contains a graph with 3 layers representing the users, roles, and resources in the organization. Each layer has nodes per user, each role, and each resource respectively. 

However, the existing graph-based policy visualization may not always be easily explainable to all administrators. Different administrators might interpret the same graphical policy representation differently, leading to different conclusions about the policies based on their level of expertise \cite{brostoff2005r}. The use of colors, line styles, and symbols used in the interface to represent rules and policies can be subject to such misinterpretations as their meanings may not be saliently described within the policy visualization interface or hidden inside separate windows \cite{morisset2018building, morisset2018visabac, bertard2020using}. For example, to get to the point where the line styles and colors used are explained in VisABAC interface, administrators have to navigate through multiple windows each time they attempt to interpret a policy while memorizing the visualization diagrams \cite{morisset2018building, morisset2018visabac}. That would negatively affect the efficiency of policy configuration \cite{maxion2005improving}. Furthermore, even if they found that information, the line styles used to denote operations "Deny Unless Permit (DUP)" and "Permit Unless Deny (PUD)" look almost similar in the interface as shown in Fig. \ref{fig: visabac}, even though they have completely opposite meanings. If administrators cannot correctly identify such subtle differences in the line styles, they might misinterpret such visualization features, resulting in an incorrect understanding of the policies.

Even though graphical policy authoring and visualization tools attempt to guide the administrator to write and visualize access control policies, it is a manual, repetitive, and laborious task that increases the administrator's overhead \cite{narouei2017identification}. This overhead can lead to fatigue and stress for the system administrator, increasing the likelihood of mistakes when configuring policies \cite{palmer_2021}. The consequences of such mistakes could become even more severe as none of the identified policy authoring and visualization tools provide adequate feedback \cite{xu2017system}, mentioning the configuration mistake (e.g., policy conflict due to an incorrectly written policy), its location (e.g., conflicting policies), the severity of the mistakes (e.g., how permissions will change if the conflicting policy is applied to the system), and solutions (e.g., how to resolve the conflict) when a mistake happens. As a result, administrators might attempt trial and error to find and resolve such mistakes, ending up adding more misconfigurations to the authorization system, causing access control failures leading to data breaches \cite{xu2017system}.
% For example, if administrators are tired and stressed after configuring policies repetitively, there is a chance that they might provide unauthorized access to a user by clicking the incorrect cell in an access matrix-based interface. 
Therefore, previous literature then proposed to remove the human factor entirely from the policy generation by proposing NLP based automated policy generation frameworks \cite{shi2011controlled, alohaly2018deep, alohaly2021hybrid, fatema2016semi, slankas2013access, heaps2021access, slankas2013accessid, brodie2006empirical, xia2022automated, alohaly2019automated, xiao2012automated, narouei2018automatic, narouei2015automatic, slankas2012classifying, singru2020efficient, vaniea2008evaluating, rosa2020parser, inglesant2008expressions, liu2017automated, narouei2017identification, ye2016identify, perry2013modular, basile2010ontology, yang2021purext, slankas2014relation, narouei2017towards, alohaly2019towards, narouei2015towards, tanoli2018towards, zhu2021unstructured, brodie2005usable, shen2021using, abdelgawad2023synthesizing}.

\subsection{NLP-based automated policy generation frameworks}
\label{subsubsec: nlpgen}

\begin{figure}[]
  \centering
    \includegraphics[height=3.8cm]{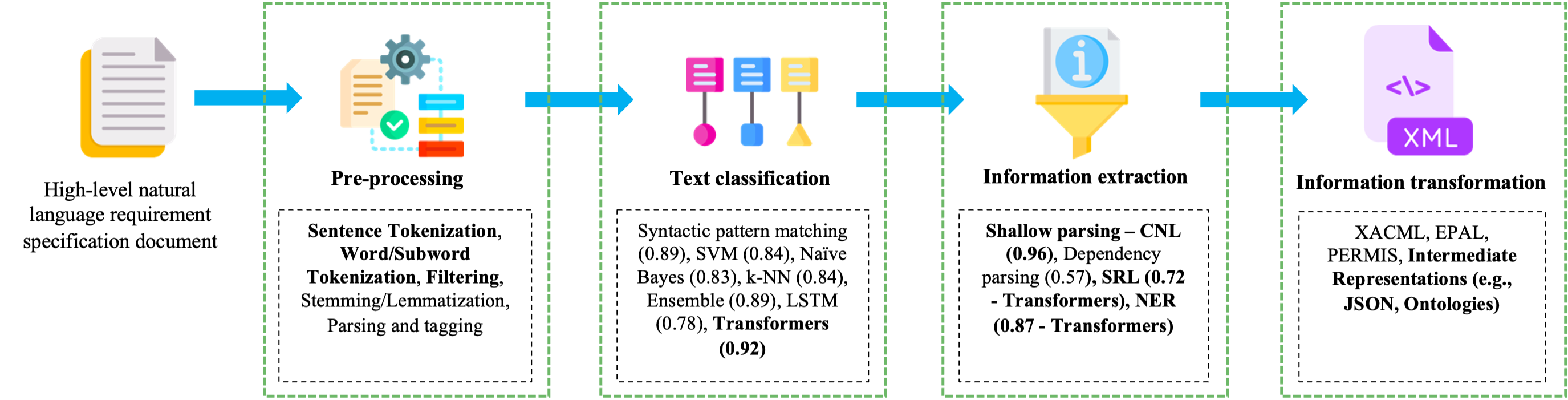}
    \caption{Automated policy generation process and the most prevalent techniques used in each step. The highest F1-score achieved using each text classification and information extraction technique is shown within brackets, highlighting the recommended technique in each step.}
    \Description[Steps of the automated policy generation process.]{The image shows the four main steps of automated policy generation namely, pre-processing, text classification, information extraction and information transformation.}
    \label{fig: pipe}
\end{figure}

The main objective of the NLP-based automated policy generation process is to translate the Natural language access control policies (NLACPs) into low-level machine executable policies with significantly low human involvement \cite{xiao2012automated, narouei2017identification, narouei2017towards}. 
% We call this translation process as "top-down policy engineering". 
It starts with the administrator's input \cite{xiao2012automated}. The input can either be a single natural language (NL) access control policy (NLACP) \cite{brodie2006empirical} or a requirement document containing multiple NL sentences \cite{narouei2015towards, xiao2012automated}. In the later input scenario, the documents are often unstructured and can contain sentences that are both NLACPs and non-NLACPs (i.e., sentences that are not access control policies) \cite{narouei2017towards}. Therefore, to achieve a successful translation, we can find four key steps followed in the previous literature., namely, \textbf{Step 1:} Pre-processing, \textbf{Step 2:} Text classification, \textbf{Step 3:} Information Extraction, \textbf{Step 4:} Information transformation, as shown in Fig. \ref{fig: pipe}.

% \begin{enumerate}

%     \item \textbf{Step 1:} Pre-processing
%     \item \textbf{Step 2:} Text classification
%     \item \textbf{Step 3:} Information Extraction
%     \item \textbf{Step 4:} Information transformation
% \end{enumerate}

By following the above four steps, first, the input NL documents were pre-processed to make them ready for translation. Then, the sentences in the pre-processed documents were classified as NLACP or non-NLACP using text classification techniques. Once the NLACPs were identified, the policy components of the NLACP (e.g., subject, action, resource, purpose, etc.) or their attributes (e.g., subject attributes, object attributes, environment attributes, etc.) were extracted. Finally, the extracted policy components or attributes were arranged as a machine-executable code to apply those policies in the authorization system. Under this subsection, we report the common NLP techniques used in the identified literature to facilitate the aforementioned steps and discuss their limitations. 
% Secondly, we report the publicly available datasets we found that can be used to train and evaluate future automated policy generation frameworks.

\subsubsection{\textbf{Step 1: Pre-processing}}
\label{subsubsec: preprocessing}
% \textbf{Step 1: Pre-processing} - 
Pre-processing is cleaning, transforming, and preparing textual data before they can be used to train or infer ML/NLP models \cite{slankas2014relation, brodie2006empirical, xiao2012automated}. 
% Even though the pre-processing techniques are essential for ML/NLP tasks, only 42.3\% () of the extracted publications explicitly mentioned the pre-processing techniques they used in the policy generation process, that utilize libraries such as CoreNLP tool kit \cite{manning2014stanford}, NLTK \cite{bird2006nltk}, and SpaCy\footnote{https://spacy.io/}. 
Some of the widely used pre-processing techniques in extracted literature are,
% The pre-processing techniques and the previous literature that employed those techniques are as follows.

\begin{itemize}
    \item \emph{Sentence Tokenization}: Sentence tokenization is the process of splitting the sentences in NL documents using punctuation marks indicating the sentence boundaries \cite{slankas2012classifying, slankas2013access, slankas2013accessid, slankas2014relation, narouei2015automatic, narouei2015towards, narouei2017identification, narouei2017towards, narouei2018automatic, tanoli2018towards, xiao2012automated, abdelgawad2023synthesizing}. This pre-processing step is necessary when processing documents containing NLACPs as followed in \cite{xiao2012automated, slankas2014relation}.

    \item \emph{Word and subword Tokenization}: Word tokenization is the process of breaking down a sentence into word tokens based on the white spaces \cite{narouei2015automatic, narouei2015towards, narouei2017identification, narouei2017towards, narouei2018automatic, alohaly2018deep, alohaly2019automated, alohaly2019towards, tanoli2018towards, yang2021purext, singru2020efficient, abdelgawad2023synthesizing, liu2017automated}. However, defining a token for each word results in a large vocabulary \cite{ho2021optimizing}. Therefore, few articles \cite{xia2022automated, heaps2021access, zhu2021unstructured} utilized subword tokenization, which operates under the principle that the rare words will be split into meaningful subwords (e.g., as a rare word, "refactoring" can be split into "re", "factor" and "ing" subwords) while others stay as word tokens \cite{ho2021optimizing}. 
    % However, when it comes to languages other than English (e.g., Chinese), rule-based tokenization was employed \cite{shen2021using}. 

    \item \emph{Filtering}: In filtering, the unwanted sections of NL requirement documents for NLACP extraction (i.e., titles, list items, headers, footers, etc.) will be removed so that only the useful parts of the input document are retained \cite{slankas2012classifying, slankas2013access, narouei2017identification, narouei2017towards}.

    \item \emph{Stemming and Lemmatization}: Stemming and Lemmatization is the process of reducing the words to their base form (e.g., liked, likes $\rightarrow$ like, caring $\rightarrow$ care) \cite{singru2020efficient, tanoli2018towards, abdelgawad2023synthesizing}.

    \item \emph{Parsing and Tagging}: Sometimes, the NL sentence was parsed and tagged according to its grammatical structure before further processing. Parsing and tagging as a pre-processing technique was performed in terms of Part-of-Speech (POS) tagging \cite{abdelgawad2023synthesizing}, shallow parsing \cite{xiao2012automated}, dependency parsing \cite{slankas2012classifying, slankas2013access, slankas2013accessid, slankas2014relation, narouei2017identification, abdelgawad2023synthesizing} and rule-based parsing \cite{perry2013modular}. 
  
\end{itemize}

Once the data were cleaned and properly formatted according to the discussed techniques, text classification techniques were applied to the resultant sentences to identify the NLACPs.

\subsubsection{\textbf{Step 2: Text Classification}}
% \textbf{Step 2: Text Classification} - 
Text classification is a NLP technique that assigns labels to textual units such as sentences, paragraphs, or documents based on certain criteria such as their sentiment or intent \cite{minaee2021deep}. In the access control policy generation process, it was mainly employed to identify NLACPs in high-level requirement specification documents \cite{xiao2012automated, slankas2012classifying, slankas2014relation, slankas2013access, slankas2013accessid}. Those documents specify how the access to company information is handled and who, under what circumstances, can access what asset \cite{narouei2017identification}. Therefore, high-level requirement specification documents often contain not only access control policies (NLACPs) of the organization but also texts that are not related to access control (non-NLACPs), requiring a text classification technique to extract access control policies to generate their machine-executable counterparts. Furthermore, text classification techniques were also employed to identify access decisions (i.e., allow or deny) \cite{heaps2021access, xia2022automated} and to identify access type (i.e., read, write, etc.) \cite{heaps2021access}. 

In the extracted publications, the text classification was done using rule-based techniques \cite{xiao2012automated, yang2021purext}, machine learning-based techniques \cite{slankas2012classifying, slankas2014relation, slankas2013access, slankas2013accessid, narouei2015automatic, narouei2015towards, narouei2017identification, narouei2018automatic, singru2020efficient}, or deep learning-based techniques \cite{alohaly2016better, narouei2015towards, narouei2017towards, alohaly2018deep, alohaly2019automated, xia2022automated, heaps2021access} as shown in Table \ref{tab: classification}. In the rule-based approach, the NLACP sentences were identified by matching the syntactic and semantic patterns of the natural language sentence with the pre-defined patterns \cite{xiao2012automated}. For instance, Xiao et al. identified the four most common grammatical patterns that define a NLACP \cite{xiao2012automated} as shown in Table \ref{tab:xiao_patterns}. Using those patterns, they filtered out the NL sentences that did not match with any of the identified patterns (i.e., sentences that do not contain access control policies). 
However, classifying sentences using a pre-defined set of patterns is limited by the number of patterns in the pattern database \cite{slankas2014relation}. If it does not contain patterns that match every possible sentence structure that can appear in NLACPs, there is a chance that actual NLACP will go unnoticed, excluding them from generating machine-executable policies. Similarly, if the NLACP is ambiguous and, in turn, does not agree with the grammatical patterns, it will also not be considered a legitimate NLACP by rule-based text classification techniques. For example, as Slankas et al. found out, only 34.4\% of the sentences in the dataset used by them agree with the four patterns identified by Xiao et al. shown in Table \ref{tab:xiao_patterns}, leaving the rest of the NLACPs undetected \cite{slankas2014relation}. These situations are particularly problematic, especially if the authorization system operates under the default-allow principle \cite{johnson2010optimizing}. For example, in default-allow systems, only those actions explicitly denied will be restricted. Suppose a policy that restricts nurses from accessing patients' personal medical records was neglected by the text classification algorithm because the policy was ambiguous or not written according to any grammatical pattern in the pattern database. In that case, since the nurses' access to personal medical records will not explicitly be restricted in the default-allow system, they will automatically gain access to those records, causing access control failures. 

Therefore, without being limited by the number of pre-defined hardcoded patterns, machine learning (ML) based algorithms were also utilized to classify NLACPs. In contrast to rule-based techniques, machine learning-based techniques learn common patterns in non-NLACP and NLACP sentences from a given training set and classify an unseen sentence based on the learned patterns \cite{slankas2014relation, slankas2012classifying, slankas2013access, slankas2013accessid}. For instance, the most prevalent ML-based classification algorithm used to identify NLACPs is the k-NN (k-Nearest Neighbours) algorithm (7 articles) \cite{slankas2014relation, slankas2012classifying, slankas2013access, slankas2013accessid, narouei2015automatic, narouei2018automatic, narouei2015towards} as shown in Table \ref{tab: classification}. k-NN classifies a given data point based on the majority vote of the existing classifications of the k nearest neighbors to the data point (i.e., the most frequent label of k data points of the training dataset closest to the given data point) \cite{slankas2012classifying}. 
% In other words, first, the k-NN algorithm calculates distances between the given query data point and every data point in the training set and sorts the distances in ascending order (i.e., closest to farthest). Then, it takes the first k distances and finds their corresponding labels from the training data. Finally, it provides the classification result for the query data point as the most frequent label in the found labels of the first k distances. 
However, finding the closest k sentences (NLACP or non-NLACP) to a given NL sentence can be tricky compared to finding the closest numerical values, as the number of attributes of each sentence is different depending on the number of words \cite{slankas2012classifying}. Therefore, Slankas et al. used a modified version of Levenshtein distance to calculate the distances between the query sentence and sentences in the training datasets \cite{slankas2012classifying, slankas2014relation}. Instead of using the number of edits needed to transform one string to another as the traditional Levenshtein distance metric does, Slankas et al. used the number of word transformations needed to convert the query sentence into a training sentence as the distance metric \cite{slankas2014relation, slankas2012classifying}. 

Apart from the traditional machine learning-based classification techniques such as Support Vector Machines (SVM) and Decision Trees, deep learning-based classification was also employed in the included literature \cite{alohaly2016better, narouei2015towards, narouei2017towards, alohaly2018deep, alohaly2019automated, xia2022automated, heaps2021access}. For instance, with the new developments in the NLP domain, transformer-based language models \cite{vaswani2017attention} were used on several occasions to identify NLACPs \cite{heaps2021access, xia2022automated} as shown in Table \ref{tab: classification}. For example, Heaps et al. \cite{heaps2021access} utilized (fine-tuned) Bi-directional Encoder Representation from Transformers (BERT) \cite{devlin2018bert} to classify user stories as NLACP, non-NLACP, and ambiguous as well as to classify their access type (i.e., read, write, etc.). 
% All the identified text classification techniques and related references are shown in Table \ref{tab: classification}.

\begin{table}[]
    \centering
    \caption{Semantic patterns in Access Control Sentences identified by Xiao et al. \cite{xiao2012automated}.}
    \label{tab:xiao_patterns}
    \begin{tabular}{ 
   >{\centering\arraybackslash}m{5cm} 
   >{\raggedright\arraybackslash}m{9.5cm}}
        \toprule
        {\textbf{Semantic pattern}} &  {\textbf{Example}}\\
        \midrule
          Modal Verb in Main Verb Group & An HCP\textsubscript{[subject]} \textbf{can view} \textsubscript{[action]} the patient’s account\textsubscript{[resource]}.\\
          
         \noalign{\vskip 3.5pt}
          Passive Voice followed by To-infinitive Phrase & An HCP\textsubscript{[subject]} \textbf{is disallowed to update} \textsubscript{[action]} patient’s account\textsubscript{[resource]}. \\
          \noalign{\vskip 3.5pt}

          Access Expression & An HCP\textsubscript{[subject]} \textbf{has read}\textsubscript{[action]} \textbf{access to} patient’s account\textsubscript{[resource]}. \\
          \noalign{\vskip 3.5pt}

          Ability Expression & An HCP\textsubscript{[subject]} \textbf{is able to read} \textsubscript{[action]} patient’s account \textsubscript{[resource]}. \\

         \bottomrule
         
    \end{tabular}
    
\end{table}

\begin{table}[]
    \centering
    \caption{Text classification techniques. Performances of each classification technique are shown in the fourth column in terms of F1 score \cite{slankas2014relation}. The full breakdown of text classification techniques can be found in \url{https://cutt.ly/gwkrEbks}.}
    \label{tab: classification}
    \begin{tabular}{ 
    >{\centering\arraybackslash}m{0.8cm} 
   >{\raggedright\arraybackslash}m{4cm} 
   >{\raggedright\arraybackslash}m{4.1cm}
   >{\raggedright\arraybackslash}m{2cm}
   >{\centering\arraybackslash}m{2.5cm}}
   
        \toprule
        & \textbf{Text classification technique} &  \textbf{Classification task} & \textbf{References} & \textbf{Highest reported performance}\\
        \midrule
        \textbf{Rule-based} & Syntactic pattern matching & NLACP classification & \cite{xiao2012automated, yang2021purext, abdelgawad2023synthesizing, shen2021using} & F1: 0.89 \cite{xiao2012automated}\\
         \midrule
         \noalign{\vskip 2.5pt}
        \multirow{5}{*}{\rotatebox[origin=c]{90}{\textbf{Machine Learning}}} & SVM (Support Vector Machines) & NLACP classification & \cite{slankas2013access, slankas2013accessid, narouei2017identification} & F1: 0.84 \cite{slankas2013access} \\
         \noalign{\vskip 3.5pt} 
         & Naive Bayes & NLACP classification & \cite{slankas2013access, narouei2017identification, slankas2013accessid} & F1: 0.83\\
          % & Policy component relation classification & \cite{slankas2014relation}\\
         \noalign{\vskip 3.5pt} 
         & k-Nearest Neighbours (k-NN) & NLACP classification & \cite{slankas2014relation, slankas2012classifying, slankas2013access, slankas2013accessid, narouei2015automatic, narouei2018automatic, narouei2015towards} & F1: 0.84 \cite{slankas2014relation}\\

         \noalign{\vskip 3.5pt}
         & Ensemble 
         % (SVM, k-NN, Naive Bayes, TF-IDF) 
         & NLACP classification & \cite{slankas2013access, slankas2013accessid, narouei2017identification} & F1: 0.89 \cite{slankas2013access}\\
         \noalign{\vskip 3.5pt} 
         & Decision Tree & Access decision classification & \cite{singru2020efficient} & F1: 0.86 \cite{singru2020efficient}\\
         \midrule
         \noalign{\vskip 2.5pt} 
         \multirow{3}{*}{\rotatebox[origin=c]{90}{\textbf{Deep Learning}}} & Convolutional Neural Networks (CNN) & Policy attribute relation classification & \cite{alohaly2018deep, alohaly2019automated} & F1: 0.85 \cite{alohaly2018deep}\\
         \noalign{\vskip 3.5pt} 
         & Long short term memory (LSTM) & NLACP classification & \cite{narouei2017towards} & F1: 0.78 \cite{narouei2017towards}\\
         \noalign{\vskip 3.5pt} 
         & Transformers (BERT) & NLACP classification & \cite{heaps2021access, xia2022automated} & F1: 0.92 \cite{xia2022automated}\\
         & & Access decision classification & \cite{xia2022automated} & F1: 0.97 \cite{xia2022automated}\\
         % \noalign{\vskip 3.5pt} 
         % Transformers (BERT) + Semantic Role Labelling (SRL) & ACP classification & \cite{xia2022automated} \\
         \bottomrule

    \end{tabular}
    
\end{table}  

However, the reliability of the ML/NLP-based text classification techniques used in existing policy generation frameworks is affected by the lack of domain-related datasets \cite{alohaly2019automated, narouei2017identification}. Domain-related datasets help the ML/NLP algorithms to be trained and adopted to the access control domain so that the model understands the patterns that are unique to access control policies. However, since there are not enough high-quality, annotated data for access control policy classification, ML/NLP models used to classify policies were not properly trained \cite{narouei2015automatic, narouei2015towards, narouei2017identification, alohaly2019automated, alohaly2019towards}. As a result, the existing policy generation frameworks sometimes might not identify actual necessary NLACPs to generate their machine-executable counterpart in the first place, resulting in missing access control policies at the end, generating security holes in the authorization system. 
% If an important policy, such as one that restricts access for a user to some resource, has gone missing, the user will automatically be granted access to the resource (in a default-allow system), causing access control failures.

% Also, by utilizing BERT-based SRL, Xia et al. classified NLACPs based on their semantic features \cite{xia2022automated}. 
% each semantic role label sequences for each predicate of the NLACP, were first pruned to remove the non-ACP verbs based semantic roles. Then, they adopt CNN to generate the word level embeddings from BERT embeddings, for each word of the sentence. Finally, they classify the sentences as ACP and non-ACP after fusing the generated embeddings and the vectorised pruned semantic role labels. 

After identifying the NLACP sentences from the NL documents, the necessary components and rules of the NL policy should be extracted to generate the machine-executable policy. Therefore, as the next step, information extraction was carried out.  

\subsubsection{\textbf{Step 3: Information Extraction}}
\label{subsubsec: ie}
% \textbf{Step 3: Information Extraction} - 
Information extraction is the process of extracting structured information from unstructured or semi-structured data sources such as NL sentences and NL documents \cite{cui2018neural}. Table \ref{tab: ie} shows the information extraction techniques used in the previous literature with the associated references.

\begin{table}[]
    \centering
    \caption{Information extraction techniques. Performances of each information extraction technique are shown in the fourth column in terms of F1 score \cite{slankas2014relation} and Accuracy (Acc.) \cite{xiao2012automated} with the used algorithm within brackets. The full breakdown of information extraction can be found in \url{https://cutt.ly/gwkrEbks}.}
    \label{tab: ie}
    \begin{tabular}{
    >{\centering\arraybackslash}m{0.6cm}
   >{\raggedright\arraybackslash}m{2.7cm} 
   >{\raggedright\arraybackslash}m{4cm}
   >{\raggedright\arraybackslash}m{2.2cm}
   >{\centering\arraybackslash}m{3.9cm}}
   
        \toprule
        & \textbf{Information extraction technique} &  \textbf{Information extraction task} &  \textbf{References} & \textbf{Highest reported performance} \\
        \midrule 
          \multirow{2}{*}{\rotatebox[origin=c]{90}{\textbf{Rule-based}}} & Syntactic parsing & NLACP component extraction & \cite{brodie2006empirical, xiao2012automated, vaniea2008evaluating, rosa2020parser, brodie2005usable, slankas2013access, narouei2017identification, ye2016identify, yang2021purext, slankas2014relation, tanoli2018towards, xia2022automated, shi2011controlled, fatema2016semi, inglesant2008expressions,turner2017proposed, shen2021using, basile2010ontology, abdelgawad2023synthesizing, liu2017automated, slankas2013accessid} & F1: 0.96 (Shallow parsing - CNL) \cite{brodie2006empirical}, 
          F1: 0.57 (Dependency parsing) \cite{slankas2014relation}\\
          \noalign{\vskip 1pt}
          & & NLACP attribute extraction & \cite{alohaly2018deep, alohaly2019automated, xia2022automated} & Not reported\\
           \midrule
          \noalign{\vskip 3.5pt}
            % \hdashline
          \multirow{4}{*}{\rotatebox[origin=c]{90}{\textbf{Deep Learning}}} & Semantic Role Labeling (SRL) & NLACP component extraction & \cite{xia2022automated, narouei2015automatic, narouei2018automatic, yang2021purext, narouei2017towards, narouei2015towards} & F1: 0.72 (Transformers) \cite{xia2022automated}\\
          \noalign{\vskip 1pt}
           & & NLACP attribute extraction & \cite{xia2022automated} & F1: 0.72 (Transformers) \cite{xia2022automated}\\
        \noalign{\vskip 3.5pt}
        % \hdashline
          & Named Entity Recognition (NER) & NLACP component extraction & \cite{heaps2021access, alohaly2019towards} & F1: 0.87 (Transformers) \cite{heaps2021access}\\
          \noalign{\vskip 1pt}
          & & NLACP attribute extraction & \cite{zhu2021unstructured} & F1: 0.80 (Transformers) \cite{zhu2021unstructured}\\
         \bottomrule
    \end{tabular}
\end{table}

\begin{figure*}[]
    \centering
    \includegraphics[height=2.5cm]{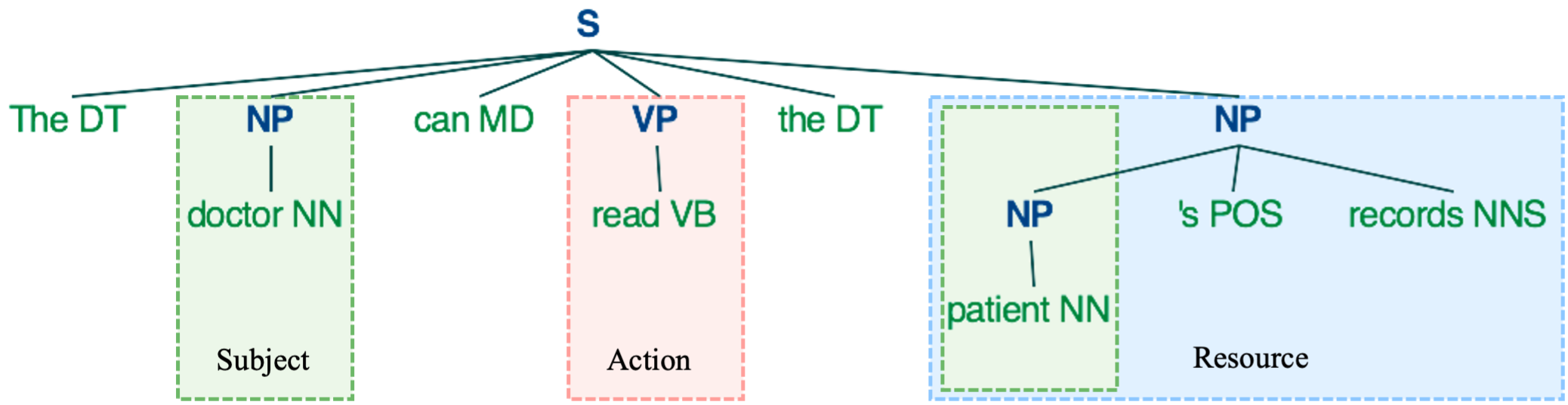}
    \caption{Shallow parse tree generated using NLTK library for the sentence \emph{"The doctor can read the patient's record"} according to the grammar rules \texttt{NP $\Rightarrow$ \{<JJ>?(<NNS> | <NN>)*\}, VP $\Rightarrow$ \{<VB>\}, NP $\Rightarrow$ \{<NP><POS><NP>\}}. NP: Noun phrase, NN: Noun, NNS: Plural noun, JJ: Adjective, VP: Verb phrase, VB: Verb in its base form, POS: Genitive marker (i.e. "'s"), MD: Modal auxiliary.}
    \Description[Shallow parse tree]{The image shows a parse tree generated by the shallow parser provided by NLTK python library. In the image subject is highlighted in green, the action is highlighted in orange, and the resource is highlighted in blue colored boxes.}
    \label{fig:shallow_parser}
\end{figure*}

According to Table \ref{tab: ie}, the most prevalent technique used to extract information from NLACPs (23 publications) is syntactic parsing. 
% performed using libraries including NLTK, SpaCy, CoreNLP, and Stanford dependency parser. 
Syntactic parsing is a method of analyzing the grammatical structure of a sentence \cite{zhang2020survey}. It identifies the syntactical relationships between words of a sentence and ultimately creates a structured representation of the sentence named "Parse tree". 
% Parse tree can further be simplified into an intermediate representation named Abstract Syntax Tree (AST), which was also used in the extracted literature to identify the policy components \cite{vanbrabant2009federated, vanbrabant2011fine}. 
There are two main syntactic parsing techniques that were widely utilized to extract access control policy components and access control rules: shallow parsing \cite{xiao2012automated, brodie2005usable, brodie2006empirical, vaniea2008evaluating, karat2006evaluating, perry2013modular} and dependency parsing \cite{alohaly2019automated, alohaly2018deep, slankas2013access, slankas2013accessid, slankas2014relation, yang2021purext, tanoli2018towards, ye2016identify, narouei2017identification, xia2022automated,abdelgawad2023synthesizing}. 
Shallow parsing, or "Chunking" divides the sentence into constituents called "Chunks" according to the syntactic structure of the natural language sentence \cite{sha2003shallow}.
% In contrast to the constituency parsing, the chunks will not be further analyzed recursively. Therefore, in shallow parsing, depth of analysis and completeness are sacrificed for the efficiency and reliability. Generating chunks is done according to a predefined set of grammar depending on the task. 
In order to form "chunks" of NLACPs, grammar rules should be developed by considering common sentence structures (i.e., templates) of NLACPs \cite{brodie2006empirical}. For example, to extract chunks, in this case, policy components in \emph{"The doctor can read the patient's record"}, the following three simple grammar rules can be used. \texttt{\textbf{Rule 1:} NP $\Rightarrow$ \{<JJ>?(<NNS> | <NN>)*\}, \textbf{Rule 2: } VP $\Rightarrow$ \{<VB>\}}, and \texttt{\textbf{Rule 3:} NP $\Rightarrow$ \{<NP><POS><NP>\}}. The above grammar rules instruct the shallow parser to create chunks in three steps. First, the shallow parser will create chunks by combining adjectives (JJ) and singular nouns (NN) or plural nouns (NNS), and tag the chunks as noun phrases (NP) according to the \texttt{Rule 1}. The affected chunks are highlighted inside green boxes in Fig. \ref{fig:shallow_parser}. Secondly, the shallow parser will tag verbs (VB) as a verb phrase (VP) according to the \texttt{Rule 2} as shown inside an orange colored box in Fig. \ref{fig:shallow_parser}. Finally, it will create chunks by combining a noun phrase (NP), a genitive marker (e.g., "'s") (POS), and another noun phrase (NP) to tag them as a noun phrase (NP) as instructed by the \texttt{Rule 3}. The generated chunks using \texttt{Rule 3} are highlighted inside a blue colored box in Fig. \ref{fig:shallow_parser}. Then the policy components such as subjects, actions, and resources can easily be extracted by identifying noun phrases that contain adjectives and nouns as the "Subjects", verb phrases as "Actions", and noun phrases built using a noun phrase, a genitive marker followed by another noun phrase as "Resources"\footnote{Modal auxiliaries (MD) and determiners (DT) will not play a role in identifying policy components.}. Following a similar procedure, Brodie et al. designed a set of grammar rules for the "SPARCLE policy workbench" that operate in order (sequentially) to add tags to the NLACPs indicating the places where policy components start and end. They used the cascaded structure to first extract the hard-to-extract components such as conditions, obligations, and purposes, and then to extract user categories, data categories, and finally actions \cite{brodie2006empirical}. However, to be able to extract components from shallow parsing, NLACPs should be written according to a specific template that was used to define grammar rules and will not be able to extract desired components otherwise. For example, if the mentioned NLACP is written in a different structure, such as \emph{"Patient's record can only be read and written by the doctor"}, the aforementioned grammar rules may not be able to parse the NLACP and identify all the necessary components. That will result in incorrect access control policies \cite{shi2011controlled}. 
% Furthermore, since the relationships between policy components cannot be identified in shallow parsing, access control rules cannot be extracted. For example, if there are multiple users performing different actions on different resources in a policy, shallow parse can only identify the users (i.e., NP), actions (i.e., VP), and resources (i.e., NP), but not which user performed what action on what resource (i.e., access control rule). 

\begin{figure*}[]
    \centering
    \includegraphics[height=2.5cm]{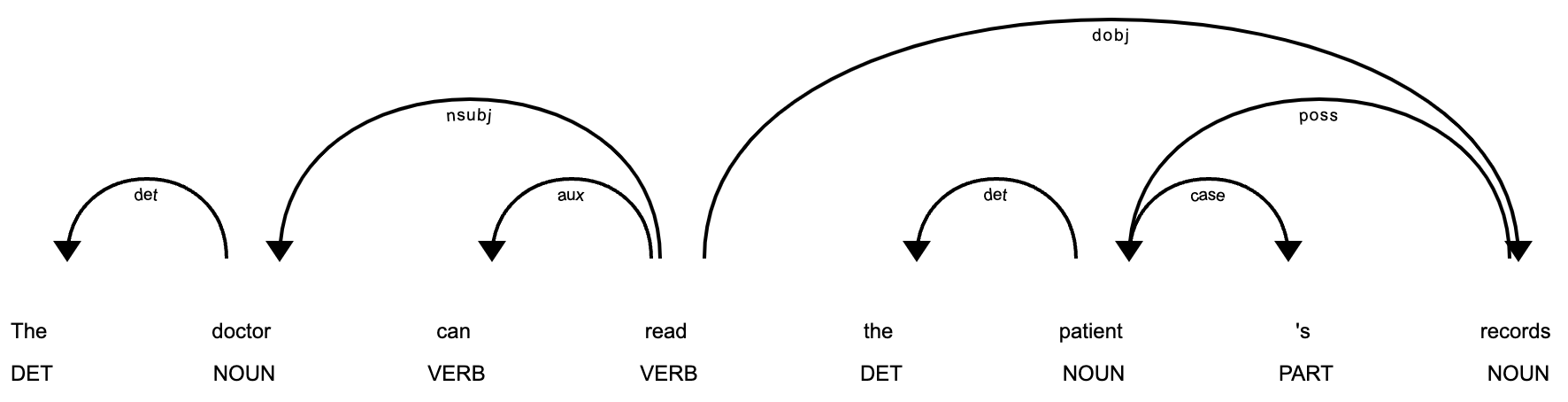}
    \caption{Dependency parse tree of the sentence \emph{"The doctor can read the patient's records"}. det: relationship between determiner (DT) and noun (NOUN), nsubj: relationship between the noun and verb (VERB), aux: relationship between auxiliary verb (AUX) and main verb, dobj: relationship between verb and object, poss: relationship between a noun and its possessive modifier, case: relationship between a noun and its case marker (PART).}
    \Description[Dependency parse tree]{The image shows the dependency parse tree generated using SpaCy python library.}
    \label{fig:dep_parser}
\end{figure*}

On the other hand, 11 of the extracted publications utilized dependency parsing to extract policy components with their relations (i.e., access control rule extraction) \cite{alohaly2019automated, alohaly2018deep, slankas2013access, slankas2013accessid, slankas2014relation, yang2021purext, tanoli2018towards, ye2016identify, narouei2017identification, xia2022automated,abdelgawad2023synthesizing}. In contrast to shallow parsing, dependency parsing identifies the relationships between words in a sentence and generates a directed graph containing the tokens as the nodes and relationships as the edges. For instance, the dependency parse tree of the NLACP \emph{"The doctor can read patient's record"} is shown in Fig. \ref{fig:dep_parser}. The relationships between word tokens are mentioned close to arrows, which cannot be extracted in shallow parsing. According to the figure, between the subject, "doctor", and the action (i.e., VERB in Fig. \ref{fig:dep_parser}), "read" has the nominal subject (nsubj) relationship. Furthermore, the direct object (dobj) relationship can be seen between the action and the resource, "records". This nsubj - VERB - dobj relationship is also the most common pattern identified by Slankas et al. according to Table \ref{tab: slankas_patterns}.
% However, the complete resource would be "patient's records". In such cases, to extract components with multiple tokens, the Depth-First Search (DFS) algorithm can be used starting from the root (i.e., the VERB "read"). For example, DFS applied to identify the resource will result in the sequence "read" $\rightarrow$ "records" $\rightarrow$ "patient" $\rightarrow$ "'s". 
Therefore, by searching for patterns that contain the nsubj and dobj relationships, access control rules were extracted in previous literature \cite{slankas2013access, slankas2013accessid, slankas2014relation, yang2021purext, tanoli2018towards, ye2016identify, narouei2017identification, xia2022automated,abdelgawad2023synthesizing}. 

To identify similar dependency patterns, Slankas et al. used the "bootstrapping" mechanism to identify different dependency relationships in NLACPs, starting from known ten seed patterns with three vertices (subject, operation, and resource) and expanded the pattern database with new patterns along the way \cite{slankas2014relation, slankas2013access}. All the known seed patterns were the same except the verb \cite{slankas2014relation}. 
 % representing ten different actions, namely create, retrieve, update, delete, edit, view, modify, enter, choose, and select. 
The subject and the resource nodes were kept as wildcards to match and extract any nouns associated with the verb. After extracting subjects and resources using the seed patterns, they expanded their pattern database in 2 ways: (1) by extracting additional dependency patterns that contain the known subjects and resources, (2) by applying a series of transformations to the existing patterns (e.g., transforming the patterns in active voice into passive voice). Some of those identified patterns in the process are listed in the 1st column of Table \ref{tab: slankas_patterns}.
% In contrast to shallow parsing, patterns identified in dependency parsing are based on the \textbf{relationships} between the words of the sentence. Thus, the dependency patterns represent the relationships between policy components including subjects, operations, and resources as well as the relationships between attributes such as subject attributes, object attributes, and environment attributes. 
In contrast to Slankas et al., Alohaly et al. manually identified the five most common relations that encode subject-attributes and object-attributes of NLACPs in \cite{alohaly2019automated} as shown in columns 2 and 3 of Table \ref{tab: slankas_patterns}. 
% In contrast to Slankas et al. \cite{slankas2014relation, slankas2013access}, Alohaly et al. manually extracted the relationship patterns from their NLACP dataset. 
Nevertheless, it is important to note that the effectiveness of using a dependency parser for policy component extraction depends on the quality and quantity of the identified patterns in the pattern database \cite{slankas2014relation}. An access control rule/attribute relation in a NLACP could not be able to extract if it does not match with a dependency pattern in the database \cite{slankas2013access}.

\begin{table}[]
    \centering
    \caption{Dependency patterns that encode relationships between the policy components (column 1), subject-attributes (column 2), and object-attributes (column 3) of Access Control Sentences. VB: verb, NN: noun, nsubj: nominal subject, dobj: direct object, nsubjpass: passive nominal subject, amod: adjectival modifier, prep: prepositional modifier, pobj: object of preposition}
    \label{tab: slankas_patterns}
    \begin{tabular}{ 
   >{\centering\arraybackslash}m{5cm} 
   >{\raggedright\arraybackslash}m{4cm}
   >{\raggedright\arraybackslash}m{3.5cm}}
        \toprule
        \textbf{Patterns between policy components \cite{slankas2014relation}} &  \textbf{Patterns between subject-attributes \cite{alohaly2019automated}} & \textbf{Patterns between object-attributes \cite{alohaly2019automated}}\\
        \midrule
          (VB root (NN nsubj) (NN dobj)) & nsubj, amod & dobj, amod\\
          
         \noalign{\vskip 2pt}
         % \hdashline

          (VB root (NN nsubjpass)) & nsubj, prep & pobj, amod \\
          \noalign{\vskip 2pt}
          % \hdashline

          (VB root (NN nsubj) (NN prep)) & nsubjpass, amod & dobj, prep\\
          \noalign{\vskip 2pt}
          % \hdashline

          (VB root (NN dobj)) & nsubj, compound & dobj, compound \\
          \noalign{\vskip 2pt}
          % \hdashline

          (VB root (NN prep\_\%)) & nsubj, ROOT, amod & nsubjpass, amod \\
          \bottomrule
    \end{tabular}
\end{table}

However, NLACPs are often ambiguous and complex \cite{narouei2015towards, inglesant2008expressions, xiao2012automated}, making it harder to parse and extract policy components correctly, especially using rule-based parsing techniques such as shallow parsing or dependency parsing \cite{shi2011controlled, brodie2006empirical}. Because those ambiguous and complex NLACPs might not agree with the parsing rules used to build parsers such as shallow parsers as it is difficult to create rules that cover all possible ambiguous and non-ambiguous NLACP sentence structures \cite{manning1999foundations}. Therefore, to avoid the ambiguities of unconstrained NLACPs, some publications have used a CNL (Controlled Natural Language) to write policies in NL, including \cite{shi2011controlled, inglesant2008expressions, brodie2005usable, brodie2006empirical, vaniea2008evaluating, shen2021using, fatema2016semi, liu2017automated}. Even though CNLs are not as strict as policy languages such as XACML, they were designed by restricting the grammar and vocabulary allowed to write policies, to prevent ambiguities, and to ensure that policies are interpreted consistently \cite{wyner2009controlled}. For example, Brodie et al. restrict the user to follow only two semantic structures to write the policy \cite{brodie2006empirical}, while Shi et al. provided five semantic structures \cite{shi2011controlled} to write policies to avoid parsing failures. Nevertheless, CNLs limit the ability to express complex policies according to their limited syntax \cite{inglesant2008expressions}, making the existing rule-based policy generation frameworks less flexible. 

% Add an example why they are less flexible in real world

Therefore, to improve the flexibility of policy generation frameworks by allowing the administrator to generate machine-executable policies from unconstrained NLACPs, deep learning-based information extraction techniques were then utilized by the previous literature \cite{xia2022automated, narouei2015towards, narouei2017towards, narouei2018automatic, yang2021purext,heaps2021access, narouei2015towards, narouei2017towards, narouei2018automatic, zhu2021unstructured}. The two most common deep learning-based techniques utilized in the extracted articles to extract information are Named Entity Recognition (NER) \cite{heaps2021access, narouei2015towards, narouei2017towards, narouei2018automatic, zhu2021unstructured} and Semantic Role Labeling (SRL) \cite{xia2022automated, narouei2015towards, narouei2017towards, narouei2018automatic, yang2021purext} as shown in Table \ref{tab: ie}. 

NER is a NLP task that identifies the named entities of a given sentence. Therefore, several extracted publications used NER to extract entities related to the access control domain, such as user, actions, resources, etc., from NLACPs \cite{heaps2021access, zhu2021unstructured}. For example, Heaps et al. fine-tuned the BERT language model \cite{devlin2018bert} using a dataset containing user stories \cite{dalpiaz2018pinpointing} to extract such named entities to build access control policies \cite{heaps2021access}. However, there is one major problem with using NER to identify policy components. Consider a NLACP, \emph{"The doctor can write the patient's record, and the nurse can only read the patient's records"}. A properly fine-tuned NER model can identify named entities, doctor and nurse as users, read and write as actions, and patient's record as a resource. However, since NER only focuses on extracting entities, it does not indicate which action is associated with which subject and what resource (i.e., access control rule). In the above example, the action \emph{"write"} belongs to the subject \emph{"doctor"} and the resource \emph{"patient's records"}. Since NER does not identify such relationships, it cannot extract access control rules in complex access control policies that contain multiple rules representing different users performing different actions on different resources \cite{heaps2021access}. As a result, it will cause access control failures, leading to data breaches. For instance, someone can generate a policy using the extracted entities from the NLACP mentioned earlier by allowing the nurse to write the patient's records since the NER output does not mention that the nurse can only read them. In such cases, NER cannot be used to generate access control policies accurately \cite{heaps2021access}.

As a solution, previous literature then utilized SRL algorithms to extract policy components that can handle multiple rule scenarios \cite{xia2022automated, narouei2015automatic, narouei2018automatic, yang2021purext, narouei2017towards, narouei2015towards}. SRL is used to analyze the meaning of the sentence by extracting its predicate-argument structure, determining "\emph{who did what to whom}", "\emph{when}", "\emph{where}", etc. \cite{shi2019simple}. Since SRL explicitly detects the subject (\emph{who}), action (\emph{what}), resource (\emph{whom}), as well as other environment attributes such as location (\emph{where}) and time (\emph{when}), previous literature extensively employed different semantic role labeling tools to extract access control rules and attributes.
% The semantic role is the relationship between a predicate (verb or the verb phrase) and the words that denote the arguments or participants of the predicate. 
Some of them are SENNA (Semantic/Syntactic Extraction using a Neural Network Architecture) \cite{collobert2011natural}, neural network-based semantic role labeler used by Narouei et al. \cite{narouei2015automatic, narouei2015towards, narouei2017towards, narouei2018automatic}, SwiRL \cite{surdeanu2005semantic} used by Narouei et al. \cite{narouei2018automatic} and Yang et al. \cite{yang2021purext}, EasySRL \cite{lewis2015joint} used by Narouei et al. \cite{narouei2018automatic}, Mate-tools Semantic Role Labeler \cite{bjorkelund2009multilingual} used by Narouei et al. \cite{narouei2018automatic} and BERT-based SRL \cite{shi2019simple} used by Xia et al. \cite{xia2022automated}.

Nevertheless, almost all the publications that used SRL for access control rule extraction did not properly adapt the used SRL algorithms to the access control domain using a domain-related dataset \cite{xia2022automated, yang2021purext, narouei2015automatic, narouei2017towards, narouei2015towards} even though adapting them to access control domain increases the access control rule extraction accuracy \cite{narouei2018automatic}. For example, Narouei et al. showed that adapting the SwiRL SRL model to the access control domain with even a small amount of labeled domain-related data, increased the rule extraction F1-score bt 2\% \cite{narouei2018automatic}. 
% They have achieved a small improvement (2\%) as there was a significant imbalance between the amount of out-of-domain data points and the amount of in-domain (i.e., access control) data points, which was about 1\% of the entire training dataset \cite{narouei2018automatic}. 
% They used a small amount of manually labeled data to re-train the SRL model via transfer learning and also utilized the pre-trained SRL model to generate pseudo labels for access control data and re-trained the model in a semi-supervised manner \cite{narouei2018automatic}. 

Instead, most of the existing works used general-purpose SRL models mentioned earlier to extract components without domain adaptation \cite{xia2022automated, yang2021purext, narouei2015automatic, narouei2017towards, narouei2015towards}. This has raised two main problems. While SRL extracts most of the required policy components, it only extracts one user and one resource for a given predicate/action of the NLACP \cite{shi2019simple}. For instance, BERT-based SRL by Shi et al. \cite{shi2019simple} identifies \emph{"The doctor and the nurse"} as a single user in the NLACP \emph{"The doctor and the nurse can read patient's records."}, despite having 2 users, \emph{"The doctor"} and \emph{"The nurse"} that belong to two rules. Therefore, to extract components with more granularity, another technique such as NER \cite{narouei2015towards, narouei2017towards, narouei2018automatic} or dependency parsing should be used on the users and resources extracted by SRL \cite{xia2022automated}. Secondly, general-purpose SRL models were often trained to generate multiple labels associated with each predicate/action in the input sentence \cite{xia2022automated}. As a result, the SRL model will extract subjects and resources associated with predicates that are not related to access control policies, such as "is", "are", etc. In the above example BERT-based SRL model \cite{shi2019simple} outputs two sequences of labels related to the two predicates, \emph{can} and \emph{read}. These additional subjects and resources related to unwanted predicates will generate redundant and incorrect access control policies, bringing the overall rule extraction accuracy down and making the maintainability of policies difficult \cite{xia2022automated}. Therefore, a pruning technique should be employed to filter the unwanted predicate-based label sequences to extract access control rules from the correct predicate-based output \cite{xia2022automated}. 

Up to this point, we discussed the techniques used in previous literature to pre-process NL documents, identify ACP sentences using text classification, and extract required policy components/rules from the identified NLACPs. As the last step, several publications then utilized information transformation formats to represent those extracted components as machine-executable codes. 

\subsubsection{\textbf{Step 4: Information Transformation}}
% \textbf{Step 4: Information Transformation} - 
% After extracting the policy components and attributes, extracted information was transformed into an intermediate representation/machine executable format in 20 of the included publications. 
The most common transformation format among those articles was XACML (eXtensible Access Control Markup Language), which was used in 7 of the identified articles \cite{ tanoli2018towards, xiao2012automated, fatema2016semi, brodie2005usable, vaniea2008evaluating, shi2011controlled, brodie2006empirical}. Apart from XACML representation of the policy, other XML-based representations such as PERMIS (PrivilEge and Role Management Infrastructure Standards) \cite{inglesant2008expressions, shi2011controlled}, and EPAL (Enterprise Privacy Authorization Language) \cite{brodie2005usable, brodie2006empirical, vaniea2008evaluating} were also employed in the extracted publications. However, if the policy generation pipeline outputs the generated policies in a specific policy language, its compatibility is reduced. For example, if the policy generation pipeline generates policies in XACML, even though the administrators need them in PERMIS language, they have to put in extra effort to translate XACML policy into a PERMIS policy. Therefore, to make the policy generation more compatible, several publications propose intermediate representations to transform generated access control policies into ontologies \cite{shi2011controlled, basile2010ontology, tanoli2018towards, ye2016identify} or JSON (JavaScript Object Notation) format \cite{rosa2020parser, liu2017automated}. 

\section{Discussion}
\label{sec:discussion}

% Due to the rapid advancement of the NLP domain, access control policy generation using natural language has also gained significant attention in recent years \cite{zhu2021unstructured, xia2022automated, heaps2021access}. 
In this systematic literature review (SLR), we analyzed \pc publications by following the guidelines proposed by Kitchenham \cite{kitchenham2004procedures} and reported  according to the PRISMA framework to identify the tools and frameworks used for access control policy configuration and generation. We reported the unique features and limitations of the previous attempts to generate access control policies from high-level natural language requirements using graphical policy authoring and visualization tools and NLP-based automated policy generation frameworks to answer the research questions RQ1 and RQ2. 

\subsection{Graphical policy authoring and visualization}
\label{subsec: discuss_gui}

Through the SLR, we revealed that the graphical policy authoring and visualization tools provide graphical interfaces that allow administrators to write and visualize policies with less cognitive load. As we reported in Section \ref{subsubsec: gui}, previous literature proposed graphical tools such as text editor-based tools \cite{stepien2009non, stepien2014non, zurko1999user, brostoff2005r}, template-based tools \cite{johnson2010optimizing, johnson2010usable, turner2017proposed, rosa2020parser}, access matrix-based tools \cite{reeder2008expandable, reeder2011more, maxion2005improving, cao2006intentional} and graph-based visualization tools \cite{morisset2018building, morisset2018visabac, bertard2020using}. These tools provide a higher level of abstraction, allowing administrators to focus on the high-level access control requirements of the organization rather than low-level technical details of the access control model, language, or syntax. 

However, despite having those advantages, we identified and discussed several limitations of the graphical policy authoring and visualization tools in Section \ref{subsubsec: gui}. 
% We discussed those limitations in terms of their lack of flexibility to support unique access requirements, inability to handle complex access control policies, and tendencies to be misinterpreted by the system administrator. 
As our SLR revealed, all those discussed limitations make the existing policy authoring and visualization tools less usable for effective access control configuration, causing access control failures \cite{reeder2008expandable, maxion2005improving, morisset2018building}. Therefore, to improve the usability of graphical policy authoring and visualization, by following Nielsen's usability guidelines \cite{nielsen1994usability}, we suggest improving the learnability, memorability, and user satisfaction of those tools while improving the accuracy and efficiency of the policy configuration process.

\begin{figure*}[]
    \centering
    \includegraphics[height=6cm]{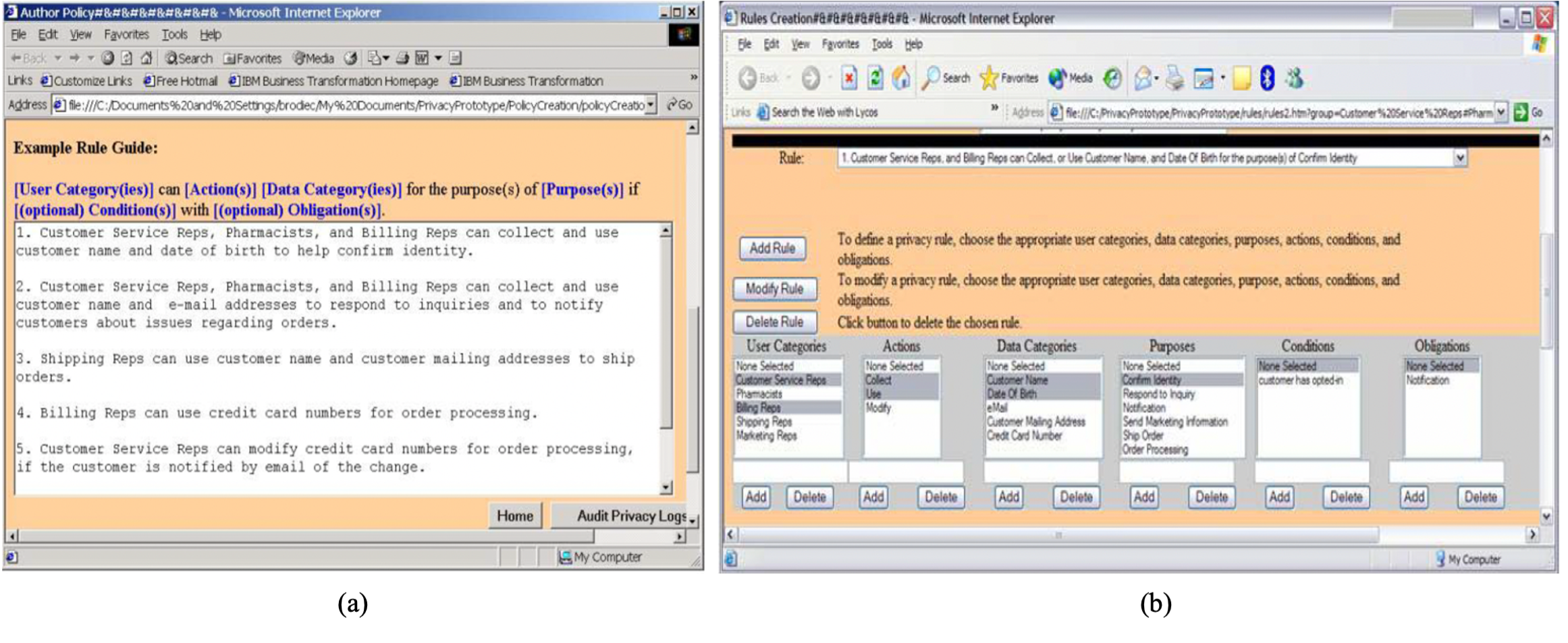}
    \caption{Different policy authoring approaches of SPARCLE Policy Workbench \cite{karat2006evaluating} that reported high user satisfaction. (a) NL with a guide approach: that provides a guide (highlighted in blue) to write access control policies in controlled natural language (CNL). (b) The Structured list approach: that allows the administrator to select policy components to build the policy as a sentence.}
    \Description[Policy authoring interfaces of SPARCLE policy workbench]{The image shows two interfaces of SPARCLE policy workbench, namely NL with a guide interface that provides a set of guidelines to write a policy, and the Structured list interface, which provides lists containing policy components to choose from, to write the policy.}
    \label{fig:satisfaction}
\end{figure*}

\emph{\textbf{Improving the learnability}} - According to Nielsen's usability components, "Learnability" measures how easy it is to perform a given policy configuration task for the first time using a policy authoring or a visualization tool \cite{nielsen1994usability}. In order to improve the learnability of those tools, they can be designed in a way that the tools are easily explainable to administrators using words, phrases, and concepts familiar to the user \cite{nielsen2005ten}, so that administrators will be able to easily interpret the functionalities of the interface and successfully configure access control policies\cite{nielsen2005ten,brostoff2005r}. Brostoff et al. utilized the mentioned learnability improvement technique to improve their policy authoring interface by simplifying its label names used to define its policy configuration features to make the interface easily explainable to administrators \cite{brostoff2005r}. In their user study, they found that the study participants were able to successfully understand the access control mechanism by referring to the labels alone, as the label names were more explainable to the participants compared to the previous versions of the interface \cite{brostoff2005r}. On the other hand, to make the policy visualization more explainable, the visualization features (e.g., colors, shapes, line styles, etc.) that were used to visualize policies, and their meanings can be described clearly in the visualization interface \cite{reeder2011more, reeder2008expandable, maxion2005improving}. To clearly describe that information, Reeder et al. \cite{reeder2008expandable, reeder2011more} have used a legend displayed at the top of their access matrix-based policy authoring and visualization interface "Expandable Grids". As a result, their interface users were able to quickly learn the interface and complete the given policy configuration tasks easily, compared to Windows XPFP (Windows XP File Permission) interface users \cite{reeder2008expandable}. Therefore, based on the mentioned empirical evidence, we suggest making the policy authoring and visualization tools easily explainable to administrators to improve their learnability.

\emph{\textbf{Improving the memorability}} - To improve the memorability of policy authoring and visualization tools, they can be designed in a way that the administrator can easily remember them and the functionality of their features \cite{nielsen1994usability}. Making the interface simple and utilizing visual cues such as icons and colors was one method followed by previous literature to make policy authoring and visualization tools memorable \cite{nielsen2005ten,stepien2009non, stepien2014non, saket2016beyond, shi2011controlled}. Stepien et al. used a simple structure to develop their text editor-based policy authoring tool containing only four text boxes to input subject, action, resource, and condition and two radio buttons to select whether the policy is an allow policy or deny policy \cite{stepien2009non, stepien2014non}. Therefore, since the interface is simple, even if administrators stop using the interface for some time, they will be able to gain the same level of proficiency in the interface quickly when they return to the interface. Furthermore, previous research found that visualization techniques, as well as colors and visual cues such as icons, help improve the memorability of the user interface significantly \cite{saket2016beyond}. For example, instead of having to read many lines of code to understand and memorize relationships between access control rules, displaying all the rules in an easily explainable visual representation such as access matrices \cite{reeder2008expandable, maxion2005improving} would help to improve the memorability of the interface. Therefore, embedding policy visualization techniques with colors and visual cues into policy authoring tools would be another way of improving the memorability of policy configuration tools.

\emph{\textbf{Improving the efficiency}} - "Efficiency" measures how quickly the administrators can perform configuration tasks once they learn the tool \cite{nielsen1994usability}. As we revealed in this SLR, one of the main reasons that prevent the administrator from efficiently configuring policies is poor representation of task-relevant information \cite{maxion2005improving}. Suppose the information relevant to configuring and understanding access control policies, such as the user's stated permissions and the user's effective permissions (i.e., permissions derived based on the user's individual permissions and permissions of the groups that the user belongs to), is either not displayed at all or hidden inside different windows of the interface. In that case, the administrator might not have a holistic idea about how the access control mechanism works and how access control rules affect one another to derive the final access decision of the policy \cite{reeder2008expandable, reeder2011more, maxion2005improving}. Therefore, if the administrator tries to write policies by searching for that information each time, the administrator's policy authoring efficiency will be decreased \cite{reeder2008expandable, maxion2005improving}. To avoid such situations, Reeder et al. \cite{reeder2008expandable} designed their interface "Expandable Grids" by displaying all the information relevant to configuring and understanding access control policies (e.g., stated permissions, effective permissions, etc.) in a single access matrix. Consequently, in the user study, Reeder et al. found that the average policy configuration task completion time of their interface users (i.e., 53.0s) is lower (by 35.3s) than the average task completion time of the Windows XPFP interface users (i.e., 88.3s) \cite{reeder2008expandable}. Therefore, we suggest displaying all the task-relevant information clearly and saliently within the interface to improve the policy configuration efficiency. 
% It is worth noting that the above suggestion will also help improve the learnability and memorability of the interface while reducing the policy configuration errors, which we discuss under "Reducing errors" next.

\emph{\textbf{Reducing errors}} - When configuring a policy, human errors can occur during four stages \cite{maxion2005improving}. Stage 1: Identify and interpret information relevant to policy configuration and decide whether or not the policy is properly configured. Stage 2: If not, formulate a sub-goal based on the interpreted information to configure the policy step by step; if the entire policy is properly configured, exit the loop. Stage 3: Formulate the plan to achieve the sub-goal. Stage 4: Execute the plan \cite{maxion2005improving}. However, if the information relevant to identifying whether the policy is correctly configured (in Stage 1) or to creating sub-goals (in Stage 2) is unavailable, incorrect, or misinterpreted, "goal errors" can occur, resulting in access control failures \cite{maxion2005improving}. The solution for those errors is to make the relevant information available to administrators in a correct, easily understandable form \cite{maxion2005improving, reeder2008expandable}. Therefore, by displaying all the information relevant to creating sub-goals in an easily understandable access matrix, Reeder et al. designed their access matrix-based policy authoring and visualization tool, "Expandable Grids" \cite{reeder2008expandable, reeder2011more}. As a result, in their user study, Reeder et al. found that the overall policy configuration accuracy of their interface users is 83.6\%, which is 27.1\% higher than the Windows XPFP users who did not have proper task-relevant information displayed in their interface \cite{reeder2008expandable}. On the other hand, even if the relevant information is available, if the interface does not support complex and unique access control policies, administrators might not be able to generate sub-goals (in Stage 2) correctly to configure complex and unique policies, leading to access control failures, as we discussed in Section \ref{subsubsec: gui}. As a solution, by following Johnson et al., we can allow administrators to write policies with different structures and policy components through the graphical policy authoring tool \cite{johnson2010usable}. 
Nonetheless, errors can still occur when applying those policies to the authorization system, such as a policy conflict in Stage 4 \cite{reeder2011more} (e.g., writing a policy by allowing a user to access a resource that is already restricted by another policy.). In such cases, administrators have to know about the exact location where the error occurred, how severe the error is, and what the possible solutions are \cite{xu2017system, nielsen2005ten}.
To do that, the policy authoring interfaces can be improved to provide feedback in a timely manner as usable error messages and warnings by emphasizing the consequences if the incorrectly written policy is applied to the system (e.g., if the conflicting policy is applied to the system, the user might gain access to confidential information of the organization.) \cite{nielsen2005ten, xu2017system}. As a result, administrators will become more cautious when writing access control policies, leading to reduced error rates \cite{nielsen2005ten}.

\emph{\textbf{Improving the subjective satisfaction}} - To make the administrator satisfied with the policy authoring experience, one technique used in previous literature is to improve the "naturalness" of the language used to write policies \cite{karat2006evaluating, inglesant2008expressions, nielsen2005ten}. By doing so, administrators were able to easily translate their mental plans into a machine-executable policy without doubting the quality of their work, leading to higher satisfaction \cite{inglesant2008expressions}. Shi et al. \cite{shi2011controlled} confirmed that theory by evaluating the satisfaction of their policy authoring interface against the traditional PERMIS policy authoring GUI through the Post-Study System Usability Questionnaire (PSSUQ) from IBM \cite{fruhling2005assessing}. PSSUQ scale ranges from 1 (no effort to use the tools) to 7 (the tool is unusable) \cite{shi2011controlled}. As Shi et al. found out, since their interface improves the "naturalness" of policy authoring compared to the traditional PERMIS GUI, their interface received the overall satisfaction score of 3.01, while the traditional GUI received the satisfaction score of 3.87 \cite{shi2011controlled}. Nevertheless, subjective satisfaction can further be improved by providing guidelines to write access control policies \cite{shi2011controlled}. To test that hypothesis, Karat et al. conducted a user study by evaluating the satisfaction of 36 policy authors when they used different policy authoring approaches \cite{karat2006evaluating}. The satisfaction was evaluated according to a questionnaire using a 7-point Likert scale (7 being the highest satisfaction) \cite{karat2006evaluating}. In that study, Karat et al. found that when the administrators were provided with either a guide to write a complete access control policy as shown in Fig. \ref{fig:satisfaction}(a) or a template to fill its blanks with the provided policy components in lists as shown in Fig. \ref{fig:satisfaction}(b), the interface achieved a higher user satisfaction (satisfaction scores of 4.9 and 4.6 respectively) compared to unguided policy authoring (satisfaction score of 3.8), which did not provide either a set of guidelines or lists of policy components \cite{karat2006evaluating}. Therefore, we suggest utilizing natural language to write policies with a clear set of guidelines on writing complete and correct access control policies to improve subjective satisfaction while improving the quality of written policies. 
     
    % However, developing a standardized access control policy configuration tool design framework to minimize the human errors of system administrators is still an open question worth exploring in future research.

\subsection{NLP-based automated policy generation}
\label{subsec: discuss_nlp}

Our SLR revealed that NLP-based automated policy generation frameworks possess the potential to generate accurate access control policies with minimum human involvement. Among many NLP techniques, previous literature employed rule-based \cite{slankas2012classifying, slankas2013access, brodie2006empirical}
% , slankas2013accessid, slankas2013implementing, slankas2014relation, xiao2012automated, brodie2005usable, brodie2006empirical, karat2006evaluating, vaniea2008evaluating, shi2011controlled, inglesant2008expressions, shen2021using, tanoli2018towards, perry2013modular, fornara2016modeling, ye2016identify, fatema2016semi, vanbrabant2009federated, vanbrabant2011fine, basile2010ontology
, machine learning-based techniques \cite{narouei2015automatic, narouei2015towards, narouei2017identification}, and deep learning-based techniques \cite{heaps2021access, xia2022automated}
% , narouei2017towards, narouei2018automatic, alohaly2018deep, alohaly2016better, alohaly2019automated, alohaly2019towards, xia2022automated, heaps2021access, zhu2021unstructured, yang2021purext, singru2020efficient, amato2013semantic
 to generate access control policies from high-level NL requirements.
% For example, as the first step of the framework, a proper dataset was selected to identify ACP patterns or evaluate the NLP models. Since high-level requirement documents can contain non-ACP sentences, only ACP sentences in the document should be identified next with the help of a classification technique such as k-NN. Once the ACP sentences were identified, policy components were extracted using an information extraction technique such as syntactic parsing. Then, the extracted components were re-arranged to generate the machine-executable access control policy in a policy language (e.g. XACML, PERMIS, etc.). Finally, the accuracy of the framework can be calculated in terms of precision, recall, or F1 score. 

However, we revealed that the existing automated policy generation frameworks are not reliable enough to generate accurate access control policies without human supervision. As we reported in Section \ref{subsubsec: nlpgen}, it is mainly because most of those frameworks inherit the limitations of the NLP techniques utilized to build those frameworks and lack domain adaptation due to the scarcity of domain-related datasets. Therefore, to develop a more reliable policy generation framework, first, it is necessary to identify what are the best (in terms of accuracy) and most prevalent techniques used by the existing automated policy generation frameworks in each step of the policy generation process: (1) pre-processing, (2) text classification, (3) information extraction, and (4) information transformation. Identifying those best techniques would help researchers to improve and combine them together to develop more reliable and secure policy generation frameworks in the future.

\begin{figure}[]
  \centering
    \includegraphics[height=9.5cm]{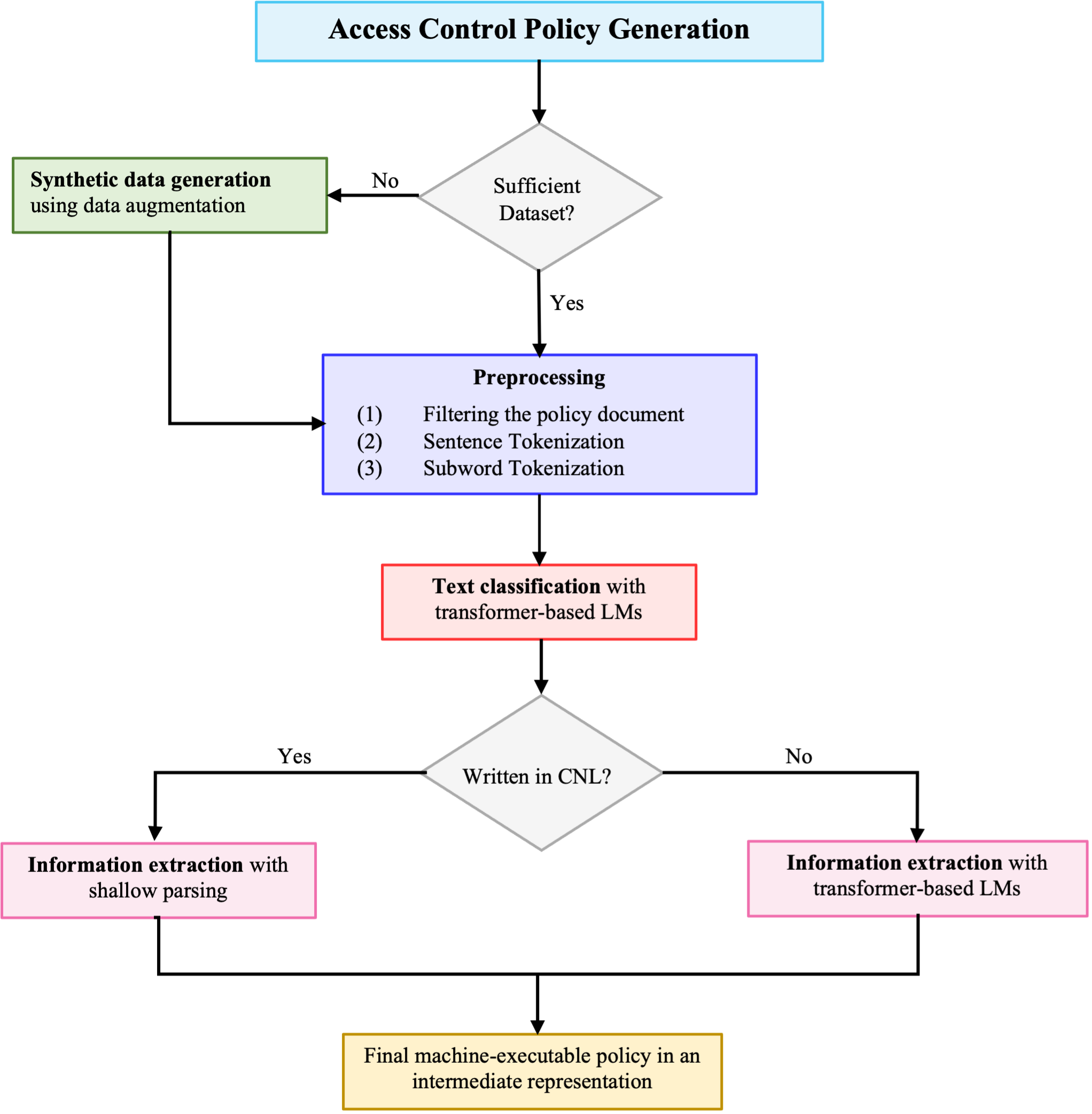}
    \caption{Framework to design reliable access control policy generation frameworks according to provided suggestions.}
    \Description[Flowchart depicting the technique selection to develop automated policy generation frameworks.]{The image shows a flow chart that describes the access control policy generation framework design. According to the flowchart, if there is sufficient data available, that data should be preprocessed first using filtering sentence tokenization, followed by sub-word tokenization first. If not, synthetic data should be generated using augmentation techniques and then carry out the pre-processing step. After pre-processing, text classification using transformer-based language models should be done to identify NLACPs of the input document. If the document was written in CNL, shallow parsing can then be used to extract policy components. If the documents are written in unconstrained NL, transformer-based LMs can be used to extract policy components. In the end, those extracted components can be used to build an intermediate representation of the NLACP, which can be easily processed by the computers.}
    \label{fig: nlp_pipeline}
\end{figure}

\emph{\textbf{Pre-processing}} - Pre-processing was carried out in the existing policy generation pipelines by using several techniques such as sentence tokenization \cite{slankas2012classifying, slankas2013access, slankas2013accessid, slankas2014relation}, word tokenization \cite{narouei2015automatic, narouei2015towards, narouei2017identification, narouei2017towards, narouei2018automatic}, stop-word removal \cite{tanoli2018towards, singru2020efficient}, and Filtering as we identified in Section \ref{subsubsec: preprocessing}. Among those techniques, filtering, sentence tokenization, and subword tokenization can be considered the most important and necessary steps to perform when generating access control policies \cite{narouei2015automatic}. After a high-level requirement specification document is obtained, first, it is necessary to filter and remove unnecessary parts of the document, such as titles, headers, footers, etc. \cite{slankas2014relation}, as we can safely assume that NLACPs will not contain within those sections. Then, the document should be tokenized to separate paragraphs, lists, etc., into sentences (i.e., sentence tokenization) and sentences into meaningful subwords (i.e., subword tokenization) \cite{slankas2014relation, narouei2015automatic, xia2022automated} to convert the sentences in the document to a set of integers based on a vocabulary (i.e., a lookup table). We recommend subword tokenization over word tokenization, as it reduces the vocabulary size, making the lookup operation faster compared to the word tokenization \cite{ho2021optimizing}. The aforementioned pre-processing steps ensure that the data used to train and infer a ML/NLP model is properly cleaned and meaningful. Therefore, we suggest the above three pre-processing steps, namely, (1) filtering, (2) sentence tokenization, and (3) subword tokenization, to perform before feeding the NL documents into a ML/NLP algorithm as shown in Fig. \ref{fig: nlp_pipeline}.

% the embedding vectors can be generated for each of the tokens to represent them in meaningful real-valued vectors to preserve their syntactic and semantic meanings using either classical \cite{alohaly2019automated, pennington2014glove} or contextual \cite{xia2022automated} pre-trained word embeddings. These four pre-processing steps ensure that the data feeding to the ML/NLP model is properly cleaned and meaningful. 
% Therefore, we suggest the above four pre-processing steps, namely, (1) filtering, (2) sentence tokenization, (3) word tokenization, and (4) embedding generation, to perform before feeding the NL documents into a ML/NLP algorithm as shown in Fig. \ref{fig: nlp_pipeline}. 

% Additionally, we suggest including "co-reference resolution" as a pre-processing technique in future policy generation research, as it will help reduce ambiguities in access control rules \cite{narouei2015automatic}. For example, in the NLACP \emph{"The doctor can read the patient's records, but he cannot write them"}, there are two access control rules that should be extracted by the ML/NLP framework: \emph{doctor can read the patient's records} and \emph{he cannot write them}. Without the first rule, there is no meaning for the second rule as \emph{"he"} and \emph{"them"} are not known. Therefore, by utilizing a co-reference resolution technique, we can modify the sentence as \emph{The doctor can read the patient's records, but \textbf{the doctor} cannot write \textbf{the patient's records}"}, resulting more meaningful rules \cite{xiao2012automated}. 

\emph{\textbf{Text classification}} - After pre-processing, pre-processed data were often fed to a text classification algorithm first to identify NLACPs \cite{slankas2012classifying, slankas2014relation, xia2022automated}. Among many text classification techniques reported in previous access control policy generation research, a transformer-based language model (LM) named BERT \cite{devlin2018bert} has achieved the highest NLACP classification performance (i.e., F1 score of 0.92 \cite{xia2022automated} for the dataset shown in Table \ref{tab: dataset}), according to Table \ref{tab: classification}. These LMs were pre-trained on gigabytes of data, enriching them with a significant understanding of NL compared to other techniques, such as rule-based parsing techniques \cite{xiao2012automated, slankas2014relation}. Therefore, those LMs are inherently better at understanding nuances of English sentences, in turn, handling ambiguous and complex NLACP structures. Thus, upon carefully adopting them to the access control domain by training them with the dataset introduced by Slankas et al. \cite{slankas2014relation}, shown in Table \ref{tab: dataset}, Xia et al. were able to produce state-of-the-art results in policy identification \cite{xia2022automated}. However, adopting a LM to access the control domain requires a relatively larger annotated dataset, since a LM often contains millions if not billions of trainable parameters to update when training \cite{devlin2018bert, xia2022automated}. Therefore, if a sufficient dataset is available, utilizing transformer-based LMs would be a promising approach for a more reliable access control policy identification. 

On the other hand, if it is difficult to collect such sufficient real-world datasets due to privacy implications, data augmentation techniques such as back translation \cite{xia2022automated}, could be used to generate more data, and annotate them manually \cite{xia2022automated} or automatically \cite{narouei2018automatic}, which we discuss later in the section. Once the dataset is expanded with more data with annotations, it can be used to fine-tune transformer-based LMs to extract access control rules, as shown in Fig. \ref{fig: nlp_pipeline}.
% if it is difficult to collect such real-world datasets due to privacy implications, using an ensemble classifier that predicts the final result based on the individual predictions of multiple relatively simple traditional ML algorithms, such as SVM, k-NN, and Naive Bayes \cite{slankas2014relation, slankas2013access}, would be the better approach according to Table \ref{tab: classification} as shown in Fig. \ref{fig: nlp_pipeline}. 

\emph{\textbf{Information extraction}} - Upon identifying NLACPs, their policy components were extracted (i.e., subject, action, resource, etc.) next. As we revealed in this SLR, the overall performance of techniques used to extract policy components depends on the language used to write NLACPs as depicted in Fig. \ref{fig: nlp_pipeline}. 
% If a diverse dataset consisting of NLACPs with different sentence structures is available, rule-based parsing techniques can be used with high reliability, irrespective of the quantity of the dataset. 
For example, if the NLACP is written in a controlled natural language (CNL) (i.e., written according to a specific template), shallow parsing \cite{brodie2006empirical} was the most promising approach (F1-score of 0.96) to extract policy components according to Table \ref{tab: ie}, as it is easy to design grammar rules for known sentence structures to achieve a higher parsing accuracy \cite{brodie2006empirical, sha2003shallow}. However, high-level requirement specification documents are often written in unconstrained natural language \cite{narouei2017identification}, which makes it difficult for shallow parsers to correctly identify policy components in them using pre-defined grammar rules as we discussed in Section \ref{subsubsec: ie}. 
% Therefore, as an alternative, dependency parsers can be used to extract access control rules (i.e., policy components and their relations together) from unconstrained natural language, as they have achieved a 0.9 parsing F1 score according to Table \ref{tab: ie}. However, all syntactic parsing techniques (i.e., shallow and dependency) are limited by either the number of grammar rules or the number of dependency patterns in the pattern database, as we identified in Section \ref{subsubsec: nlpgen}. 
Therefore, in that case, according to previous literature, we suggest utilizing transformer-based LMs to extract access control rules when developing a policy generation framework in the future. Because according to Table \ref{tab: ie}, they were able to achieve F1 scores of 0.87 in extracting policy components using NER when there is only one rule in the policy and 0.72 in extracting access control policy components as meaningful rules via SRL when there are multiple rules in a policy respectively. 

% However, F1-score of 0.72 indicates that the information extraction technique does not accurately extract policy components every time, resulting in incorrectly identified or missed policy components sometimes. Those occasions result in incorrect access control policies that might lead to data breaches. Therefore, as a solution, we suggest utilizing the human-in-the-loop technique \cite{wu2022survey} in a productive way to utilize the administrator's expertise to refine the incorrect model outputs when necessary.

\emph{\textbf{Information transformation}} - Choosing the policy language that can represent extracted access control rules depends on the type of policies that the organization uses \cite{narouei2015automatic}. For example, according to previous literature, if the organization is using ABAC (Attribute-based Access Control) policies, the recommended language would be XACML as it is specifically designed to represent ABAC policies \cite{brodie2006empirical, shi2011controlled, xia2022automated, alohaly2019automated}. On the other hand, previous literature recommended PERMIS language when they are dealing with RBAC (Role-based Access Control) policies \cite{brostoff2005r, inglesant2008expressions}. However, if the generated policies are in a specific policy language such as XACML, they might not be compatible with an organization that uses PERMIS and vice versa. Therefore, by keeping compatibility, we suggest generating the final access control policies in an intermediate representation such as an ontology \cite{shi2011controlled, inglesant2008expressions, basile2010ontology, tanoli2018towards} or JSON format \cite{rosa2020parser}, which can be easily processed and extract rules to generate any machine-executable policies in any policy language. 

\subsubsection{\textbf{Datasets}} - According to the above discussion, one of the key factors that decide what technique to use to identify NLACPs and extract their policy components/rules accurately is the availability of datasets. However, as we revealed in this SLR, many previous studies have highlighted that the access control policy engineering domain suffers from a scarcity of domain-related data \cite{narouei2017identification, xia2022automated}. Nevertheless, we came across one dataset which is widely used among the extracted literature, including \cite{narouei2015automatic, narouei2017towards, alohaly2016better,alohaly2018deep,alohaly2019automated,alohaly2019towards}, introduced by Slankas et al. in \cite{slankas2014relation}. The dataset consists of five data sources containing 2477 sentences from multiple real-world systems, such as iTrust \cite{meneely2012appendix}, IBM course registration system, CyberChair, and the Collected ACP data from \cite{xiao2012automated}. Detailed information about the dataset is shown in Table \ref{tab: dataset}, with the highest F1 scores achieved for each data source in the Text Classification (TC F1) and Information Extraction (IE F1) steps of the access control policy generation process.

\begin{table}[]
    \centering
    \caption{Statistics of the dataset compiled by Slankas et al. \cite{slankas2014relation} and highest F1 scores achieved for each dataset in Text Classification (TC F1) and Information Extraction (IE F1).}
    \label{tab: dataset}
    \begin{tabular}{ 
   >{\centering\arraybackslash}m{3.6cm} 
   >{\centering\arraybackslash}m{1.5cm}
   >{\centering\arraybackslash}m{2cm}
   >{\centering\arraybackslash}m{2cm} | 
   >{\centering\arraybackslash}m{1.6cm}
   >{\centering\arraybackslash}m{1.6cm}}
        \toprule
        \textbf{Data source} &  \textbf{Total sentences} & \textbf{ACP sentences} &  \textbf{non-ACP sentences} & \textbf{TC F1} & \textbf{IE F1}\\
        \midrule
          iTrust for Text2Policy \cite{xiao2012automated} & 471 & 418 & 53 & 0.98 \cite{narouei2017identification, slankas2014relation} & 0.8 \cite{narouei2018automatic}\\
          iTrust for ACRE \cite{slankas2014relation} & 1160 & 549 & 611 & 0.9 \cite{narouei2017identification} & 0.72 \cite{xia2022automated}\\
          % \noalign{\vskip 2pt}
          IBM Course Management & 401 & 168 & 233 & 0.97 \cite{xiao2012automated} & 0.72 \cite{xia2022automated}\\
          % \noalign{\vskip 2pt}
          CyberChair & 303 & 140 & 163 & 0.79 \cite{xia2022automated} & 0.71 \cite{xia2022automated}\\
          CollectedACP \cite{xiao2012automated} & 142 & 114 & 28 & 0.92 \cite{narouei2017identification} & 0.82 \cite{narouei2018automatic}\\
          % \noalign{\vskip 2pt}
          \midrule
          \textbf{Total sentences} & 2477 & 1389 & 1088 & - & -\\
          % \noalign{\vskip 2pt}
          \textbf{Proportions} & - & 56\% & 44\% & - & -\\
          \bottomrule
    \end{tabular}
\end{table}

However, the above dataset is not large and diverse enough to train a transformer-based LM \cite{xia2022automated}. Therefore, previous literature used data augmentation techniques such as back translation, which translates a sentence from the dataset into a different language and translates it back to the original language, 
to generate more synthetic data points from the existing data \cite{xia2022automated}. Once the dataset is expanded with synthetic data, it should be annotated to train a model in a supervised manner \cite{xia2022automated}. The annotation process can be done mainly in two ways: manual \cite{xia2022automated, narouei2018automatic} and automated \cite{narouei2018automatic}. In the manual annotation process, experienced human annotators were used to generate labels for the dataset manually \cite{xia2022automated, narouei2018automatic}. 
% Once the dataset is fully annotated, the NLP models can be trained (i.e., domain adaptation \cite{narouei2018automatic}) to identify NLACPs and extract their rules.
However, manual labeling is laborious, expensive, and time-consuming \cite{narouei2018automatic}. Therefore, Narouei et al. used a semi-supervised learning technique named "pseudo labeling" to automatically generate pseudo labels for the unlabeled data using a pre-trained SRL model, SwiRL \cite{narouei2018automatic}. Then, the pseudo-labeled small in-domain dataset was mixed with a large out-of-domain dataset and re-trained the model to achieve a 2\% increment in F1 score in access control rule extraction \cite{narouei2018automatic}. By using the aforementioned techniques, sufficient and annotated datasets can be created to adapt NLP models to generate access control policies with higher reliability. 
% Furthermore, the bootstraping technique that Slankas et al. used to identify dependency pateterns of the NLACP dataset can also be considered as a automated labelling technique as it automatically identify the patterns and

\section{Limitations}
\label{sec:threats}

This SLR was conducted thoroughly to provide an extensive overview of the topic by preserving the reproducibility of the reported results in the literature. Nevertheless, while conducting the SLR, as we first filter the returned articles from the search query based on their titles and abstracts alone using our inclusion and exclusion criteria, a relevant article may be excluded during the selection phase. Therefore, to avoid such situations as much as possible, we performed an additional manual search, a backward snowballing search to include publications cited by the publication extracted in digital library and conference/journal search phases \cite{desolda2021human}. Then, we thematically analyzed the publications \cite{braun2012thematic} to answer the research questions mentioned in Section \ref{subsubsec:rqs}. The first author performed the analysis in a systematic way to generate codes and identify themes (patterns) that help answer the research questions. However, the generated codes and themes can be biased depending on the experience, knowledge, and point of view of the coder. To reduce this bias, as Braun and Clerk advised \cite{braun2012thematic}, we considered the perspectives of all the authors when developing themes, as we mentioned in Section \ref{subsubsec: data_analysis}.

% the consistency of the coding process was checked by involving a co-author to validate the descriptions of the themes and data coding. We conducted three meetings to analyze the generated codes and themes, and the disagreements were discussed. As a result, the differences between the co-authors were resolved.  

% Also, we set database alerts for the original search query to extract articles published between the publication retrieval and final documentation. Additionally, publications that report negative and inconclusive results are rare in the retrieved articles, as they may be less likely to be published. Thus, publication bias may exist, which could lead to an overestimation of the capabilities of the GUI-based tools and NLP-based frameworks. 

% \textbf{Study selection} - The review may be subjected to selection bias as some studies such as the studies that are indirectly related to the policy generation may be excluded during the study selection process. To avoid such cases we conducted a rigorous screening process according to pre-defined inclusion and exclusion criteria. Also, the inclusion and exclusion criteria were discussed and refined in weekly meetings with the co-authors to reduce the subjective bias, so that no relevant study is overlooked in the selection phase \cite{blanco2022software}.  % Provide an exmaple

\section{Conclusion and Future works}
\label{sec:conclusion}

Access control failures due to usability and reliability issues of the existing policy configuration tools and generation frameworks could lead to data breaches \cite{xu2017system, inglesant2008expressions}. Therefore, to improve their usability and reliability, we conducted a SLR analyzing (1) graphical policy authoring and visualization tools and (2) NLP-based automated policy generation frameworks to reveal their limitations. Based on our findings, we have provided several design guidelines that would help improve the usability of policy authoring and visualization tools according to Nielsen's usability components \cite{nielsen1994usability}: learnability, memorability, efficiency, errors, and subjective satisfaction. On the other hand, we further provided guidelines to improve the reliability of automated policy generation frameworks by selecting the best techniques for each step of the policy generation process: (1) pre-processing, (2) text classification, (3) information extraction, and (4) information transformation, as well as developing and annotating datasets. Next, based on the research gaps revealed through the SLR, we will highlight several future works that would help address the usability-security trade-off of access control policy generation approaches in the future.

According to our SLR, incorporating the administrator's perspective via user studies to develop policy authoring and visualization tools may improve their usability \cite{brostoff2005r, morisset2018building}. In the light of the experienced administrators might prefer textual interfaces such as Command Line Interfaces (CLIs) as they like the control that textual interfaces provide compared to graphical interfaces \cite{botta2007towards}. On the other hand, less experienced administrators might prefer GUIs as they allow the administrator to write (i.e., policy authoring) and visualize policies without worrying about access control languages and syntax \cite{botta2007towards, stepien2009non}. Therefore, due to those differences, including administrators with different levels of expertise in the tool design process via user studies will help create more usable tools for experienced and inexperienced administrators alike in the future. 

After designing and developing graphical policy authoring and visualization tools, they should be evaluated to verify their usability \cite{nielsen1994usability} with the involvement of human subjects/participants via usability evaluation instruments such as PSSUQ (Post-Study System Usability Questionnaire) \cite{fruhling2005assessing}, or System Usability Scale (SUS) \cite{brooke1996sus}. However, existing studies rarely used those standard usability evaluation instruments to evaluate policy authoring and visualization tools and refine them accordingly \cite{stepien2009non, stepien2014non, brodie2005usable}. Those instruments provide standard questionnaires to evaluate the usability of user interfaces in terms of the discussed usability components in Section \ref{subsec: discuss_gui} \cite{fruhling2005assessing, brooke1996sus}. Therefore, if access control policy authoring and visualization tools were not evaluated using those questionnaires, there is a chance those tools to be unusable, leading the administrators to make mistakes when writing and interpreting policies, as we revealed in Section \ref{subsec: discuss_gui}. For example, if the tool developers did not collect feedback from user study participants on how easy it is to learn the tool (which is one of the main items of those usability questionnaires) and refine the tool accordingly, the tool might be harder to learn, resulting in mistakes when configuring access control policies. As a result, administrators might tend to misinterpret the functionalities of the interface, producing incorrect access control policies that lead to data breaches \cite{brostoff2005r,reeder2008expandable, bauer2009real}. Thus, utilizing standard usability evaluation instruments \cite{fruhling2005assessing} such as PSSUQ and SUS to evaluate the usability of access control policy configuration tools can be done as future work. 

However, those existing instruments might not correctly evaluate the usability of access control configuration tools. Because as we identified in this SLR, their usability also depends on factors such as support for complex access requirements, avoiding misinterpretations of policy visualizations, and the ability to identify and resolve policy conflicts \cite{reeder2007usability}, which are not explicitly covered in the general usability evaluation instruments such as PSSUQ and SUS \cite{reeder2007usability}. Therefore, we encourage researchers to develop standard usability evaluation instruments to specifically evaluate access control policy authoring and visualization tools in the future.

An important item in PSSUQ is "The system gave error messages that clearly told me how to fix problems.", which evaluates feedback from the system \cite{fruhling2005assessing}. 
However, as we discussed in Section \ref{subsubsec: gui}, existing policy authoring tools do not provide sufficient feedback by pointing out policy authoring mistakes, their locations, their severities, and how to resolve the mistakes \cite{xu2017system, nielsen2005ten}. If such feedback is not provided clearly, as Xu et al. found out, administrators often try trial and error to find and correct the mistakes, sometimes introducing more errors to the authorization system \cite{xu2017system}. 
Therefore, to provide such feedback, first, that feedback should be carefully designed in a precise and concise way that is easily understandable by highlighting the severity of mistakes \cite{nielsen2005ten}, irrespective of the administrator's expertise. To do that, Explainable Security (XSec) concepts \cite{vigano2020explainable} can be used to decide what information should be presented in the feedback, where and when to display the feedback, and how the feedback should be displayed while emphasizing its severity. Therefore, we suggest conducting research on designing feedback (i.e., error messages, warnings \cite{nielsen2005ten}) with the help of XSec specifically for access control policy configuration systems as future works. 

Once the feedback is designed, it should be generated based on the written policy automatically. Therefore, future research can then focus on improving the existing automated policy generation frameworks to automatically generate feedback/insights on the poorly written (i.e., ambiguous, incomplete, etc.) NLACP, by taking the administrator's expert level into account. By doing so, administrators will become more cautious when writing policies, and they do not need to use trial and error to find solutions, leading to fewer access control failures. 

However, to adapt ML/NLP techniques to generate insights on poorly written NLACPs or even to generate machine-executable policies, diverse, correctly annotated datasets are required, as we discussed in Section \ref{subsec: discuss_nlp}. 
Hence, developing such datasets should also be done as a part of future research, which will help accurately generate machine-executable policies and feedback on poorly written NLACPs automatically.
% , which can be immensely helpful for the further development of access control policy engineering research. 
Once datasets are developed, NLP/ML models such as transformer-based LMs can be trained using techniques such as transfer learning \cite{weiss2016survey} or Parameter Efficient Fine Tuning (PEFT) \cite{houlsby2019parameter} to generate feedback on the poorly written NLACPs. 
% When the feedback is generated, it should be presented to the administrator in a precise and concise way that is easily understandable by highlighting the severity of mistakes \cite{nielsen2005ten}, irrespective of their expertise. To do that, Explainable Security (XSec) concepts \cite{vigano2020explainable} can be used to decide where and when to display the feedback and how the feedback should be displayed while emphasizing its severity. Therefore, then, we suggest conducting research on designing feedback (i.e., error messages, warnings \cite{nielsen2005ten}) with the help of XSec specifically for access control policy configuration systems. 

Furthermore, as we revealed in this SLR, automated policy generation frameworks might not be 100\% reliable even if the most advanced NLP techniques that are proven to provide more accurate results in general text classification and information extraction tasks such as transformer-based LMs \cite{devlin2018bert, raffel2020exploring, liu2019roberta, ouyang2022training} are used to build them \cite{xia2022automated, heaps2021access}. As a solution, while the policy generation framework provides feedback to the administrator, the administrator's expertise can also be utilized to provide feedback on incorrectly generated policies by the policy generation frameworks via a usable interface. That feedback can be used to re-train the underlying policy generation framework with Reinforcement Learning with Human Feedback (RLHF) \cite{ouyang2022training} to improve the automated access control policy generation framework further. Adapting these techniques to optimize the automated policy generation framework (especially text classification and information extraction steps) and combining it with a usable interface that supports the aforementioned feedback mechanisms can be another important future research direction. 

That combined framework will help avoid data breaches due to access control failures in two ways. First, since it improves the administrator's policy authoring experience via a usable interface and a usable feedback mechanism that provides insights on poorly written policies, human mistakes that lead to data breaches will be alleviated. Secondly, since it uses advanced NLP techniques to improve the reliability of the underlying policy generation process with human (i.e., administrator) feedback, errors of automated policy generation will also be reduced, leading to fewer access control failures in the future.

\bibliographystyle{ACM-Reference-Format}
\bibliography{sample-base}


\end{document}